\documentclass[aps,prl,twocolumn,superscriptaddress,groupedaddress,reprint]{revtex4} 

	\usepackage{amsmath}
	\usepackage{bbm}
	\usepackage{makeidx}
	\usepackage{amsfonts}
	\usepackage[ansinew]{inputenc}
	\usepackage[usenames,dvipsnames]{pstricks}
	\usepackage{subfigure}
	\usepackage{epsfig}
	\usepackage{pst-grad} 
	\usepackage{pst-plot} 
	\usepackage[colorlinks,hyperindex]{hyperref}
	\usepackage{mathrsfs} 
	\usepackage{braket}
	\hypersetup
	{
		colorlinks,%
		citecolor=black,%
		linkcolor=black,%
		urlcolor=black,%
	}

	\newcommand*\diff{\mathop{}\!\mathrm{d}}

	\newcommand{\appropto}{\mathrel{\vcenter{
  \offinterlineskip\halign{\hfil$##$\cr
    \propto\cr\noalign{\kern2pt}\sim\cr\noalign{\kern-2pt}}}}}
	
\makeindex

\usepackage{rotating} 

\begin{document}

\title{Destabilization of local minima in analog spin systems by correction of amplitude heterogeneity}

\author{Timoth\'ee Leleu}
\affiliation{Institute of Industrial Science, The University of Tokyo, 4-6-1 Komaba, Meguro-ku, Tokyo 153-8505, Japan}
\author{Yoshihisa Yamamoto}
\affiliation{ImPACT program, The Japan Science and Technology Agency, Gobancho 7, Chiyoda-ku, Tokyo 102-0076, Japan,}
\affiliation{E. L. Ginzton Laboratory, Stanford University, Stanford, CA94305, USA.}
\author{Peter L. McMahon}
\affiliation{E. L. Ginzton Laboratory, Stanford University, Stanford, CA94305, USA.}
\affiliation{National Institute of Informatics, 2-1-2 Hitotsubashi, Chiyoda-ku, Tokyo 101-8430, Japan.}
\affiliation{School of Applied and Engineering Physics, Cornell University, Ithaca, NY14853, USA}
\author{Kazuyuki Aihara}
\affiliation{Institute of Industrial Science, The University of Tokyo, 4-6-1 Komaba, Meguro-ku, Tokyo 153-8505, Japan}
\affiliation{International Research Center for Neurointelligence (IRCN), The University of Tokyo, 7-3-1 Hongo, Bunkyo-ku, Tokyo 113-0033, Japan.}



\begin{abstract}
The relaxation of binary spins to analog values has been the subject of much debate in the field of statistical physics, neural networks, and more recently quantum computing, notably because the benefits of using an analog state for finding lower energy spin configurations are usually offset by the negative impact of the improper mapping of the energy function that results from the relaxation. We show that it is possible to destabilize trapping sets of analog states that correspond to local minima of the binary spin Hamiltonian by extending the phase space to include error signals that correct amplitude inhomogeneity of the analog spin states and controlling the divergence of their velocity. Performance of the proposed analog spin system in finding lower energy states is competitive against state-of-the-art heuristics.
\end{abstract}

\maketitle

Many algorithms and hardware dedicated to solving hard combinatorial optimization problems utilize a mapping of the cost function to the energy landscape of simple physical systems such as classical spins \cite{Kirkpatrick1983,Wang2015,Zarand2002,Boettcher2001,Sutton2017}, quantum spins \cite{Kadowaki1998,Farhi2000}, optical oscillators \cite{Wang2013,Inagaki2016,Marandi2014,McMahon2016,Clements2017}, solid-state oscillators\cite{Wang2017,Csaba2018}, and neural networks \cite{Hopfield1985} (see \cite{Percus2006,Mezard2009,Lucas2014} for reviews). From a random initial state, the probability to find the lowest energy states, or ground states, depends critically on whether the non-equilibrium dynamics of such systems can escape efficiently from local minima. The difficulty in solving hard problems stems from the fact that the number of these local minima, for which energy is not decreased by any single spin flip at zero temperature, generally grows exponentially with the problem size. By coupling the system to a (Markovian) heat bath, transitions over energy barriers are allowed, and convergence to the ground state is assured for a slow enough decrease of the temperature \cite{Kirkpatrick1983,Geman1984}. Such approaches have been improved upon in recent years \cite{Wang2015}, and generalized to the case of quantum spins in quantum annealing \cite{Kadowaki1998,Farhi2000}.

Alternatively, it has been suggested in various systems (including soft spins \cite{1981Sompolinsky}, analog neural networks \cite{Hopfield1985}, coherent Ising machines \cite{Wang2013,Marandi2014,McMahon2016,Inagaki2016}) that relaxing the binary spins $\sigma_i=\pm 1$ to analog values $x_i$ with $x_i \in \mathbb{R}$ may increase the probability of finding lower energy states. In these gain-dissipative systems, combinatorial optimization is classically achieved by mapping local minima to fixed point attractors\cite{Hopfield1985,Wang2013}. These attractors are created because of dissipation that induces phase-space contraction under the action of the dynamics\cite{Strogatz1996}. By tuning the gain, i.e., the average energy supplied to the system, dissipation can be compensated and the rate and directions in which phase-space volumes are contracted can be controlled. Earlier studies on the Hopfield neural network suggest that reduction of the gain (or steepness of the neuron transfer function) results in improved quality of solutions of traveling salesman problems \cite{Hopfield1985} and in the exponential decrease of the number of fixed points \cite{Fukai1990,Waugh1990}. Moreover, gradual reduction of the gain, which can be related to temperature via the naive mean field of Thouless, Anderson, and Palmer, can serve as ``mean field annealing'' \cite{Bilbro1989,Peterson1989}. In the framework of the coherent Ising machine, it has been shown numerically \cite{Wang2013} and experimentally \cite{Marandi2014,McMahon2016} that such machines can be used as an efficient heuristic solver for hard combinatorial optimization problems such as MAX-CUT (which is equivalent to Ising problems). These schemes rely on setting the gain to a minimum at which (most) suboptimal configurations, or excited states, cannot be stable fixed points\cite{Wang2013,Leleu2017}. In the field of combinatorial optimization, such analog systems have been relatively less studied than their binary spins counterparts, and it is believed that they do not perform as well as state-of-the-art heuristics \cite{Wilson1988} notably because it is difficult to map low energy configurations of the binary system to analog states with smaller loss\cite{Baum1986,Leleu2017}, although these systems are ideal for efficient implementation on dedicated hardware \cite{Hopfield1985,Marandi2014,Schuman2017}.

In this Letter, we argue that analog bistable systems, even when simulated on a classical computer, can in fact find low energy states at least as efficiently as current state-of-the-art heuristics. Our model is based on the observation that the improper mapping of the objective function by the loss landscape results from the fact that the amplitudes of analog variables are in general heterogenous (i.e., not all equal). We propose to correct the amplitude heterogeneity by extending the phase space \cite{Sourlas2005} using auxiliary degrees of freedom, which we call error variables. The proposed system utilizes the fact that the local stability (the Jacobian matrix) of fixed points can be controlled \textit{a priori} in analog systems in order to reduce the number of stable local minima. Moreover, we show that periodic and chaotic attractors can be avoided by controlling the gain such that the divergence of the error signal velocity is close to zero but positive, which ensures that phase-space volumes in the auxiliary subspace never contract and, in turn, forbids the creation of attractors. The error variables play the role of a non-Markovian reservoir that guarantees positive entropy production in the system despite dissipation. We have performed numerical simulations demonstrating that this effect generalizes well by finding the ground states of spin glasses and solving Ising problems from standard benchmark sets against state-of-the-art heuristics, and suggest that orders of magnitude decrease of the time-to-solution can be obtained in the case of an implementation on an analog Ising machine.

We consider a network of analog bistable units $x_i$, $i \in \{1,\ldots, N\}$, for which time-evolution is given as follows:

\begin{align}\label{eq:softspins}
\frac{\diff x_i}{\diff t} = f_i = \phi(x_i) + e_i I_i,
\end{align}

\noindent where $\phi(x_i)$ represents the time-evolution of isolated units with $\phi(x_i)=-\frac{\partial V_b}{\partial x_i}$ and $V_b = -(-1+p) \frac{x_i^2}{2} + \frac{x_i^4}{4}$ the paradigmatic bistable potential. The function $\phi(x_i)$ can be written as $\phi(x_i)=-x_i + p x_i - x_i^3$ in which the first, second, and third terms can be interpreted as the loss, linear gain with rate $p$, and nonlinear saturation. These dynamics can be used to describe various systems such as soft spins \cite{1981Sompolinsky}, open-dissipative quantum systems such as degenerate parametric oscillators \cite{Leleu2017,Shoji2017,Yamamura2017,Wustmann2013}, and weakly coupled neural networks near pitchfork bifurcations \cite{Hoppensteadt1997}. For solving the combinatorial optimization problem that is defined by the cost function $V(\boldsymbol{\sigma})$, the coupling $I_i$ is chosen as $I_i \propto - \frac{\partial V (\boldsymbol{x}) }{\partial x_i}$, i.e., $I_i$ is the gradient of the potential $V$. In particular, Ising problems with the cost function $V(\boldsymbol{\sigma})=\mathcal{H} = -\frac{1}{2} \sum_{ij} \omega_{ij} \sigma_i \sigma_j$ can be solved using the injection term $I_i = \epsilon \sum_j \omega_{ij} x_j$ with $\epsilon$ the coupling strength ($\epsilon > 0$, $\omega_{ii}=0$, and $\omega_{ij}=\omega_{ji}$ for $j \neq i$). 

In order to destabilize states that correspond to local minima of the Ising Hamiltonian, we propose to control the target amplitude, noted $a$ with $a>0$, of the variables $x_i$ independently of the linear gain $p$ by considering the following error signal induced by amplitude heterogeneity:

\begin{align} \label{eq:error}
\frac{\diff e_i}{\diff t} = g_i = - \beta (x_i^2-a) e_i,
\end{align}

\noindent where $e_i$ and $\beta$ are the error variables and the rate of change of error variables, respectively, with $e_i>0$ and $\beta>0$\footnote{Note that exponentially increasing error variables have been previously proposed in \cite{Ercsey-Ravaz2011} for solving constraint satisfaction problems.}. 

First, we examine the existence of fixed point attractors. The fixed points of the dynamical system described by eqs. (\ref{eq:softspins}) and (\ref{eq:error}) are given as follows:

\begin{align} \label{eq:homoa}
\begin{cases}
\frac{\diff e_i}{\diff t} = 0,\\
\frac{\diff x_i}{\diff t} = 0,
\end{cases}
\implies 
\begin{cases}
x_i^2 = a,\\
 e_i = \frac{1-p+a}{\epsilon h_i \sigma_i},\end{cases}
\forall i,
\end{align}

\noindent where $h_i$ are the elements of the vector $\boldsymbol{h} = \Omega \boldsymbol{\sigma}$ and $\Omega$ the matrix of couplings with $\Omega=\{\omega_{ij}\}_{ij}$. The configuration $\boldsymbol{\sigma}$ corresponds to the sign of the state $\boldsymbol{x}$ at the fixed point. Note that the analog states $x_i$ are exactly binary at the steady state with $x_i = \sigma_i \sqrt{a}$. Moreover, the internal fields $h_i$ are such that $h_i \sigma_i > 0$, $\forall i$, at equilibrium when $p<1$ because $e_i >0$, $\forall i$. Thus, all fixed points of the analog system correspond to local minima of the binary spin system at $T=0$.

The linear stability of these fixed points can be examined by analyzing the following Jacobian matrix:

\begin{align}\label{eq:Jgeneral}
J = \left[ {
\begin{array}{cc}
    J_{xx} & J_{xe} \\
    J_{ex} & J_{ee} \\
\end{array}}
\right],
\end{align}

\noindent with $J_{xx} = (-1+p-3a) I + \epsilon D[\boldsymbol{e}] \Omega$, $J_{xe} = \epsilon \sqrt{a} D[\boldsymbol{h}]$, $J_{ex} = - 2 \beta \sqrt{a} D[\boldsymbol{\sigma} \cdot \boldsymbol{e}]$, and $J_{ee} = 0$. Moreover, $D[\boldsymbol{X}]$ is the diagonal matrix with elements $D_{ii}[\boldsymbol{X}] = X_i$ and $D_{ij}[\boldsymbol{X}] = 0$ for $i \neq j$, where $X_i$ expresses the components of a vector such as $e_i$ or $h_i$. Note that $J_{ex} J_{xe} = b I$, with $b = -2 \beta a (1-p+a)$. The eigenvalues $\lambda_j^{\pm}$ of the Jacobian matrix can be explicitly calculated by considering its characteristic polynomial, and are given as follows:

\begin{align}
\lambda_j^{\pm} =
\begin{cases}
\frac{1}{2}(-2a + (1-p+a) \mu_j \pm \sqrt{\Delta_j}) \text{ if } \Delta_j > 0,\\
\frac{1}{2}(-2a + (1-p+a) \mu_j \pm i \sqrt{-\Delta_j}) \text{ otherwise},
\end{cases}
\end{align}

\noindent with $\Delta_j = (-2a + (1-p+a) \mu_j)^2 + 4b$, $i^2=-1$, and $\mu_j$ the $j^{th}$ eigenvalues of the matrix $\tilde{\Omega} = D[(\boldsymbol{\sigma} \cdot \boldsymbol{h})^{-1}] \Omega - I$ with $(\boldsymbol{\sigma} \cdot \boldsymbol{h})^{-1} = \{(h_i \sigma_i)^{-1}\}_i$. Because the vector $\boldsymbol{\sigma} \cdot \boldsymbol{h}$ has positive components at local minima, the eigenvalues of $D[(\boldsymbol{\sigma} \cdot \boldsymbol{h})^{-1}] \Omega$ are the same as the ones of $D[(\boldsymbol{\sigma} \cdot \boldsymbol{h})^{-1}]^{\frac{1}{2}} \Omega D[(\boldsymbol{\sigma} \cdot \boldsymbol{h})^{-1}]^{\frac{1}{2}}$ (Sylvester's law of inertia), which is a symmetric real matrix. Thus, the eigenvalues $\mu_j$ are always real.

The $2N$ eigenvalues of the system are always pairs $\lambda_j^{+}$ and $\lambda_j^{-}$ with $j \in \{1,\ldots,N\}$. Each pair become the same real value when $\Delta_j = 0$, i.e., under the condition given as follows:

\begin{align}
\theta = G_{\pm}(\beta,\mu_j),
\end{align}

\noindent with $\theta=\frac{a}{1-p}$ and $G_{\pm}(\beta,\mu) = \frac{4 \beta - \mu (\mu-2) \pm 4 \sqrt{\beta (\beta + \mu)}}{(\mu-2)^2 - 8 \beta}$.

\begin{figure}[htp]
\centerline{\includegraphics[width=0.55\textwidth]{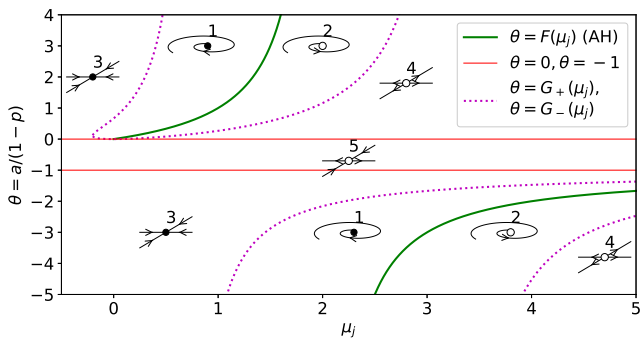}}
\caption{\label{fig:1} Bifurcation diagram in the space $\{\theta = \frac{a}{1-p}, \mu_j (\boldsymbol{\sigma}) \}$ at configuration $\boldsymbol{\sigma}$. Thick lines correspond to the sets where $\text{Re}[\lambda_j^{\pm}] = 0$, i.e., $\theta = 0$, $-1$, or $F(\mu_j)$ with $\Delta_j<0$, and dotted lines to the sets where $\theta = G_{\pm}(\beta,\mu_j)$. In the regions marked 1 and 2, the pair of eigenvalues $\lambda_j^{+}$ and $\lambda_j^{-}$ are complex conjugate with negative real parts and positive real parts, respectively, whereas in regions 3, 4, and 5, they are both real with both, none, and one of their real parts being negative, respectively. $\beta=0.2$. AH: Andronov-Hopf bifurcation.}
\end{figure}

The stability of fixed points depends on the sign of the real part of the eigenvalues of the Jacobian matrix. It can be shown that $\text{Re}[\lambda_j^{+}] = 0 $ in the following set (note that $\text{Re}[\lambda_j^{+}] \geq \text{Re}[\lambda_j^{-}]$, $\forall j$):

\begin{align} \label{eq:stabcond}
\text{Re}[\lambda_j^{+}] = 0 \Leftrightarrow
\begin{cases}
\theta=0 \text{ or } \theta=-1, \text{ if } \Delta_j>0,\\
\theta= F(\mu_j (\boldsymbol{\sigma})), \text{ if } \Delta_j<0,
\end{cases}
\end{align}

\noindent with $F(\mu) = -\frac{\mu}{(\mu-2)}$. The parameter $\theta$ can be interpreted as the ratio between the target amplitude $a$ and the effective loss (the loss minus the linear gain $p$) of the analog system. The stability of fixed points is illustrated in Fig. \ref{fig:1} in the space $\{\theta = \frac{a}{1-p}, \mu_j(\boldsymbol{\sigma}) \}$ where $\theta$ depend on the controllable parameters of the analog system whereas $\mu_j(\boldsymbol{\sigma})$ are determined by the spin configuration $\boldsymbol{\sigma}$ and couplings $\Omega$. We denote $\mu_0(\boldsymbol{\sigma})$ the maximum eigenvalue of $\tilde{\Omega}$ calculated at the fixed point $\boldsymbol{x}$ corresponding to the configuration $\boldsymbol{\sigma}$. The state $\boldsymbol{x}$ becomes unstable at a supercritical Andronov-Hopf bifurcation when the real part of the dominant eigenvalue of the Jacobian matrix J, noted $\lambda_0(\boldsymbol{\sigma})$, becomes positive, i.e., $\mu_0(\boldsymbol{\sigma}) > F^{-1}(\theta)$ for $p<1$ (see eq. \ref{eq:stabcond}).

Next, we consider the existence of limit cycles and chaotic attractors. A necessary condition for the existence of these attractors is that the divergence of the velocity, or equivalently, the rate of change of phase-space volumes, is negative at the proximity of the trapping sets\cite{Strogatz1996}. In order to estimate the divergence at configuration $\boldsymbol{\sigma}$, we utilize the fact that the fast subsystem on $\boldsymbol{x}$ converges to the slow manifold characterized by $\frac{\diff x_i}{\diff t} \approx 0$, $\forall i$, in the limit $\beta \ll 1$. Moreover, the analog states become approximately binary in the limit of small coupling strength ($\epsilon \ll 1$) with $x_i^2$ given as $x_i^2 = (x_i^{(0)})^2 + 2 \epsilon x_i^{(0)} x_i^{(1)} + \mathcal{O}(\epsilon^2)$ with $(x_i^{(0)})^2 = -1+p$ and $2 \epsilon x_i^{(0)} x_i^{(1)} = \epsilon e^{(0)} h_i \sigma_i$ for $p>1$ (see supplementary material). The term $e^{(0)}(t)$ is the zeroth order approximation of $e_i(t)$ with $e_i(t) = e^{(0)}(t) + \epsilon e_i^{(1)}(t) + \mathcal{O}(\epsilon^2)$, which is valid for $e^{(0)}(t) \epsilon \ll -1+p$, and $e^{(0)}(t) = e^{- \beta \int (-1+p-a) \diff t' }$ for $e_i(0) = 1$, $\forall i$. Moreover, the first order term $e_i^{(1)}$ is given as $e_i^{(1)} = -\beta \epsilon e^{(0)} [\int_0^t h_i(t') \sigma_i(t') \diff t']$ and depends on the previous history of states visited. Although this Taylor approximation is valid only for $p>1$, numerical simulations show that it is still accurate by continuity for $p<1$ and $|1-p| \ll 1$ for a finite value of $\epsilon$. Then, the divergence of the vector field $\boldsymbol{g}$, defined as $\text{div } \boldsymbol{g} = \sum_i \frac{\partial g_i}{\partial e_i}$ can thus be approximated as follows (see supplementary material):

\begin{align}
\text{div } \boldsymbol{g} = \beta [N(1-p+a) + 2 \epsilon e^{0}(t) \mathcal{H}(t) + \mathcal{O}(\frac{(e^{(0)})^2 \epsilon^2}{p-1}) ] \label{eq:divapprox2}.
\end{align}

\begin{figure*}[htp]
\centerline{\includegraphics[width=1.1\textwidth]{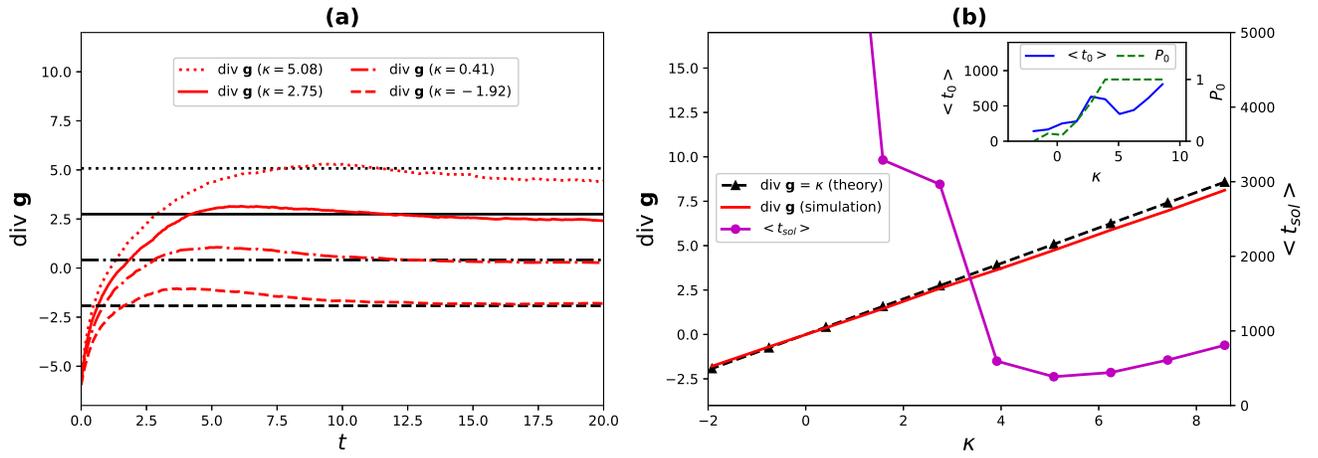}}
\caption{\label{fig:2} Control of the divergence in the auxiliary subspace. (a) Divergence $\text{div } \boldsymbol{g}$ in the auxiliary subspace vs. normalized simulation time $t$. After some transient behavior, the divergence is constant and approximately described by its analytical value $\text{div } \boldsymbol{g} \approx \kappa$. (b) Divergence $\text{div } \boldsymbol{g}$ and average time $<t_{sol}>$ for finding the ground state of the Ising Hamiltonian with a probability superior to $99\%$ with $t_{sol} = <t_0> \frac{log(0.1)}{log(1-P_0)}$ for $P_0 < 0.99$, and $t_{sol} = <t_0>$ otherwise, with $P_0$ and $<t_0>$ (see inset) the probability of finding the ground state during a single run and averaged time-to-solution for successful runs, respectively.}
\end{figure*}

\begin{figure}[htp]
\centerline{\includegraphics[width=0.5\textwidth]{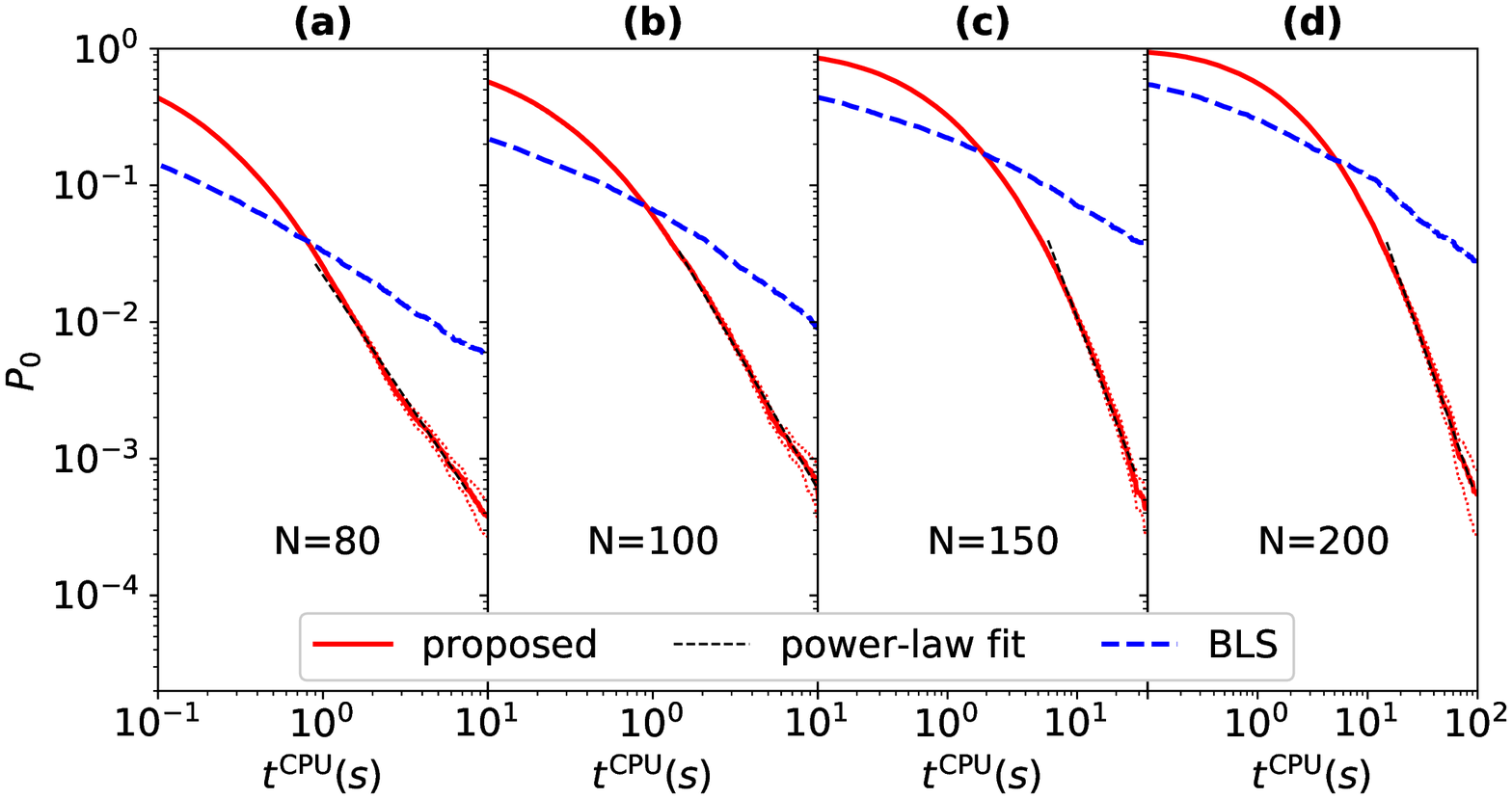}}
\caption{\label{fig:3} Proportion $p_0(t)$ of instances unsolved, i.e., for which a ground-state configuration has not been found, after the CPU time $t$ in seconds of the SK problems of size $N=80$, $N=100$, $N=150$, and $N=200$ shown in (a), (b), (c), and (d), respectively, for the proposed scheme and the breakout local search (BLS) algorithm\cite{Benlic2013} in the logarithmic scale. The $95\%$ confidence interval shown by dotted lines is calculated using 20 runs per instance.}
\end{figure}

In order to prevent the system from being trapped in limit cycles and chaotic attractors, we propose to set the divergence of velocity in the auxiliary subspace on $\boldsymbol{e}$ to be always positive along the trajectories of the system. This can be achieved by modulating the target amplitude $a$ as follows:

\begin{align}
a &= \alpha + \epsilon <e_i(t_c) h_i(t) \sigma_i(t)> \label{eq:mod2}, 
\end{align}

\noindent where $\alpha$ is the target amplitude baseline and $t_c$, with $t_c<t$, is the time of the last change of configuration $\boldsymbol{\sigma}$, i.e., the time at which one of the $x_i$ has changed its sign. Moreover, $<X_i>$ denotes the ensemble average of $X_i$ with $<X_i> = \frac{1}{N}\sum_i X_i$. The control scheme described in eq. (\ref{eq:mod2}) implies, using eq. (\ref{eq:divapprox2}), that the divergence is approximately given as $\text{div } \boldsymbol{g} \approx \kappa$ with $\kappa = N \beta (1 - p + \alpha)$ as long as the configuration $\boldsymbol{\sigma}(t)$ switches rapidly before the error variables $\boldsymbol{e}$ blow up. The parameter $\kappa$ can be set to $\kappa >0$, $\forall t$, by modulating $a$ and $p$ such that the two conditions given in eqs. (\ref{eq:stabcond}) and (\ref{eq:mod2}) are respected, which ensures that there are no stable fixed points at suboptimal configurations and that the divergence in the auxiliary subspace is always positive, respectively. Note that, when $\beta$ is very small, the fast variables $\boldsymbol{x}$ are slaved to the slow error signals $\boldsymbol{e}$. Thus, positive divergence in the auxiliary subspace should imply that the whole system cannot be trapped in a region of negative divergence at the proximity of an attractor. In order to verify this claim, we have simulated the proposed system using the weights $\omega_{ij}$ from 5000, 5000, 3000, and 2000 instances of Sherrington-Kirkpatrick (SK) spin-glass problems (an NP-hard problem) of size $N=80$, $N=100$, $N=150$, and $N=200$ spins, respectively\footnote{See the supplementary material for more details about the simulation scheme.}. For an instance of size $N=100$, Figure \ref{fig:2} (a) shows that the system exhibits constant phase-space volume expansion or contraction in the auxiliary subspace $\boldsymbol{e}$ when $\kappa >0$ and $\kappa <0$, respectively. Moreover, the system does not become trapped in an undesirable attractor and the ground-state configuration of the Ising Hamiltonian is visited with probability $P_0 = 1$ from any random initial condition when $\kappa \gg 0$ (see Figs. \ref{fig:2} (b) and \ref{fig:3}). As the constant divergence in the auxiliary subspace is increased, the system always visits a ground-state configuration, but the time-to-solution increases, which is likely a consequence of the fact that the dynamics becomes more complex\cite{Ercsey-Ravaz2011}. Therefore, the optimal regime for finding lower energy states is for positive but small divergence of the auxiliary subspace velocity. In this case, the probability of finding the ground state from a randomly chosen set of initial conditions is superior to 99.9 percent even for $N=200$ (see Fig. \ref{fig:3}). Thus, numerical simulations suggest that the system does not become bounded within a subspace. Further analysis is necessary, however, for clarifying the conditions, such as the maximal value of $\beta$, under which this is the case. The proposed dynamics finds ground-state configurations more reliably than the state-of-the-art algorithm, called BLS\cite{Benlic2013}, for harder problem instances. Figure \ref{fig:3} shows that the proportion of unsolved instances $p_0(t)$ is well fitted by a power law such that $p_0 \sim t^{-\gamma(N)}$ for $t \gg 0$. The positive divergence implies that error variables eventually become very large. However, the parameter $\kappa$ can always be chosen sufficiently small such that a ground-state configuration is visited before the error variables blow up, which is confirmed by our numerical simulations. In practice, the error variables are rescaled such that $e_i(t) = \frac{e_i(t^-)}{<e_i(t^-)>}$, with $t^-<t$, whenever the mean error signal $<e_i(t^-)>$ is superior to the threshold, noted $\Gamma$, in order to insure the stability of numerical simulations. Note that the proposed scheme does not guarantee that the best solution found after $t$ is the optimal one (but see \cite{Molnar2018} for a heuristic approach to predicting the likelihood of optimality for MaxSAT problems).

The increase of dynamical complexity that results from the addition of the amplitude-heterogeneity error correction scheme can be interpreted in terms of entropy. With the adiabatic elimination of the fast variables $\boldsymbol{x}$, the sign of the entropy production rate $\frac{\diff S}{\diff t}$\cite{Andrey1985,Ruelle2008}, defined as $\frac{\diff S}{\diff t} = \int \rho(\boldsymbol{e}) \text{div } \boldsymbol{g} \diff \boldsymbol{e}$ with $S$ the Gibbs entropy $S = - \int \rho(\boldsymbol{e}) ln \rho(\boldsymbol{e}) \diff \boldsymbol{e}$ and $\rho(\boldsymbol{e})$ the probability density of states, is equal to the sign of $\kappa$ when using the control of divergence proposed in eq. (\ref{eq:mod2}). Thus, the coupling to the auxiliary subspace implies that the system has always a positive production of entropy despite dissipation when $\kappa >0$.

For further comparison against state-of-the-art algorithms, the proposed scheme is adapted for solving MAX-CUT problems of the G-set \cite{Benlic2013}, which is a reference benchmark used in the community of combinatorial optimization. When simulated on a desktop computer, the scheme can achieve performance in terms of solution quality and time-to-solution that is qualitatively similar\footnote{Note that computer simulations are obtained using CPUs with similar clock rates (Xeon E5440 2.83 GHz and Xeon E5-2680 2.70 GHz for BLS, a Xeon X5690 3.47 GHz for the proposed scheme), but using different programming languages (C++ for BLS and Matlab for the proposed scheme). Therefore, the averaged time-to-solutions are presented as indicative purposes for comparison of their orders of magnitudes.} to that of BLS\cite{Benlic2013}, which itself outperforms other recent heuristics on MAX-CUT problems. For several instances, the proposed scheme finds solutions of better quality than previously known from \cite{Benlic2013} (see the supplementary material for the details of the benchmark on the G-set). Importantly, the advantage of the proposed scheme relies on the ability to implement it on hardware dedicated to solving Ising problems in the analog domain \cite{Yamamoto2017}. If we assume that the presented scheme is realized in hardware that has the same experimental parameters as existing coherent-Ising-machine implementations \footnote{Estimation of the time-to-solution $<t^{\text{CIM}}>$ in the case of an implementation on the coherent Ising machine is calculated using the normalized time-to-solution $<t>$ obtained from numerical simulations of the proposed scheme and the value of the time normalization constant (the cavity photon lifeftime) that has been obtained experimentally \cite{Marandi2014,McMahon2016,Inagaki2016}, i.e., $<t^{\text{CIM}}> \approx 10^{-5} <t>$. This approximation is justified by the fact that the computational effort related to the amplitude heterogeneity error correction scheme (see eqs. (\ref{eq:error}) and (\ref{eq:mod2})) is negligible compared to the one related to the coherent Ising machine (see eq. (\ref{eq:softspins})).}, then we predict that such a hardware solver will feature a time-to-solution for solving Ising problems that is a factor of 100-1000 smaller than that of the state-of-the-art classical heuristic algorithms running on a conventional desktop computer, even when the problem size is not large, i.e., $N \ll 1000$.

This research was supported by ImPACT Program of Council for Science, Technology and Innovation (Cabinet Office, Government of Japan). We thank Ryan Hamerly and the anonymous reviewers for comments on the manuscript. We were informed of related work about amplitude heterogeneity error correction (see \cite{Kalinin2018}) during the preparation of this manuscript.

\clearpage

\bibliographystyle{apsrev4-1}

%

\pagebreak
\onecolumngrid

\setcounter{equation}{0}
\setcounter{figure}{0}
\setcounter{table}{0}
\setcounter{page}{1}
\renewcommand{\theequation}{s\arabic{equation}}
\renewcommand{\thefigure}{s\arabic{figure}}
\renewcommand{\bibnumfmt}[1]{[s#1]}
\renewcommand{\citenumfont}[1]{s#1}

\section*{Supplementary material}

\subsection{Approximation of the state-space velocity divergence in the limit $\beta \ll 1$ and $\epsilon \ll 1$}

The time-evolution of the proposed system, described using eqs. (1) and (2), can be rewritten as follows:

\begin{align}
\frac{\diff x_i }{\diff t}  &= (-1+p) x_i - x_i^3 + \epsilon e_i \sum_j \omega_{ij} x_j, \nonumber \tag{s1}
\end{align}

\noindent with $e_i(t)$ given as $e_i(t) = e_i(0) e^{-\beta \int_0^t (x_i(t')^2 - a(t')) \diff t'}$ after integrating eq. (2). Then, the state $x_i(t)$ of the system can be approximated at the proximity of the steady-state $\frac{\diff x_i }{\diff t} = 0$ using the Taylor approximation $x_i = x_i^{(0)} + \epsilon x_i^{(1)} + \mathcal{O}(\epsilon^2)$ when $p>1$. At the zeroth order, the state can be described as follows: 

\begin{align}
(x^{(0)}_i)^2 &= (-1+p) \nonumber \tag{s2},\\
e^{(0)}_i &= e_i(0) e^{-\beta \int_0^t (-1+p(t')-a(t')) \diff t'}. \nonumber \tag{s3}
\end{align}

\noindent In the following, $e^{(0)}_i$ is simply written $e^{(0)}$ because it is independent of the index $i$ when $e_i(0)=1$, $\forall i$. At the first order, the state is described as follows:

\begin{align}
x^{(1)}_i &= \frac{e^{(0)} \sum_j \omega_{ij} x^{(0)}_j}{2 (x^{(0)}_i)^2},\nonumber \tag{s4} \\
e^{(1)}_i &= - \beta e^{(0)} \int_0^t 2 x^{(0)}_i(t') x^{(1)}_i(t') \diff t'.\nonumber \tag{s5}
\end{align}

Moreover, the divergence of state-space velocity $\boldsymbol{f}$ and $\boldsymbol{g}$ are given as follows (see eqs. (1) and (2)):

\begin{align}
\text{div } \boldsymbol{f} &= \sum_i \frac{\partial f_i}{\partial x_i} = N(-1+p) - 3 \sum_i x_i^2,\nonumber \tag{s6}\\
\text{div } \boldsymbol{g} &= \sum_i \frac{\partial g_i}{\partial e_i} = -\beta (\sum_i x_i^2 - a N).\nonumber \tag{s7}
\end{align}

Using the Taylor approximation of $x_i^2$, equal to $x_i^2 = (x^{(0)}_i)^2 + 2 \epsilon x^{(0)}_i x^{(1)}_i + \mathcal{O}(\epsilon^2)$, the divergence is given as follows:

\begin{align}
\text{div } \boldsymbol{f} &= -2N(-1+p) + 6 \epsilon e^{(0)}(t) \mathcal{H} + \mathcal{O}(\frac{(e^{(0)})^2 \epsilon^2}{p-1}),\nonumber \tag{s8}\\
\text{div } \boldsymbol{g} &= \beta [ N(1-p+a) + 2 \epsilon e^{(0)}(t) \mathcal{H} + \mathcal{O}(\frac{(e^{(0)})^2 \epsilon^2}{p-1})].\nonumber \tag{s9}
\end{align}

Although the Taylor approximations given in eqs. (s2) to (s9) are only valid for $p>1$, numerical simulations show that these are still accurate by continuity for $p<1$ and $|1-p| \ll 1$ for a finite value of $\epsilon$ (see supplementary Fig. \ref{fig:s1} in the case $e_i = 1$, $\forall i$).

\begin{figure*}[htp]
\centerline{\includegraphics[width=1.0\textwidth]{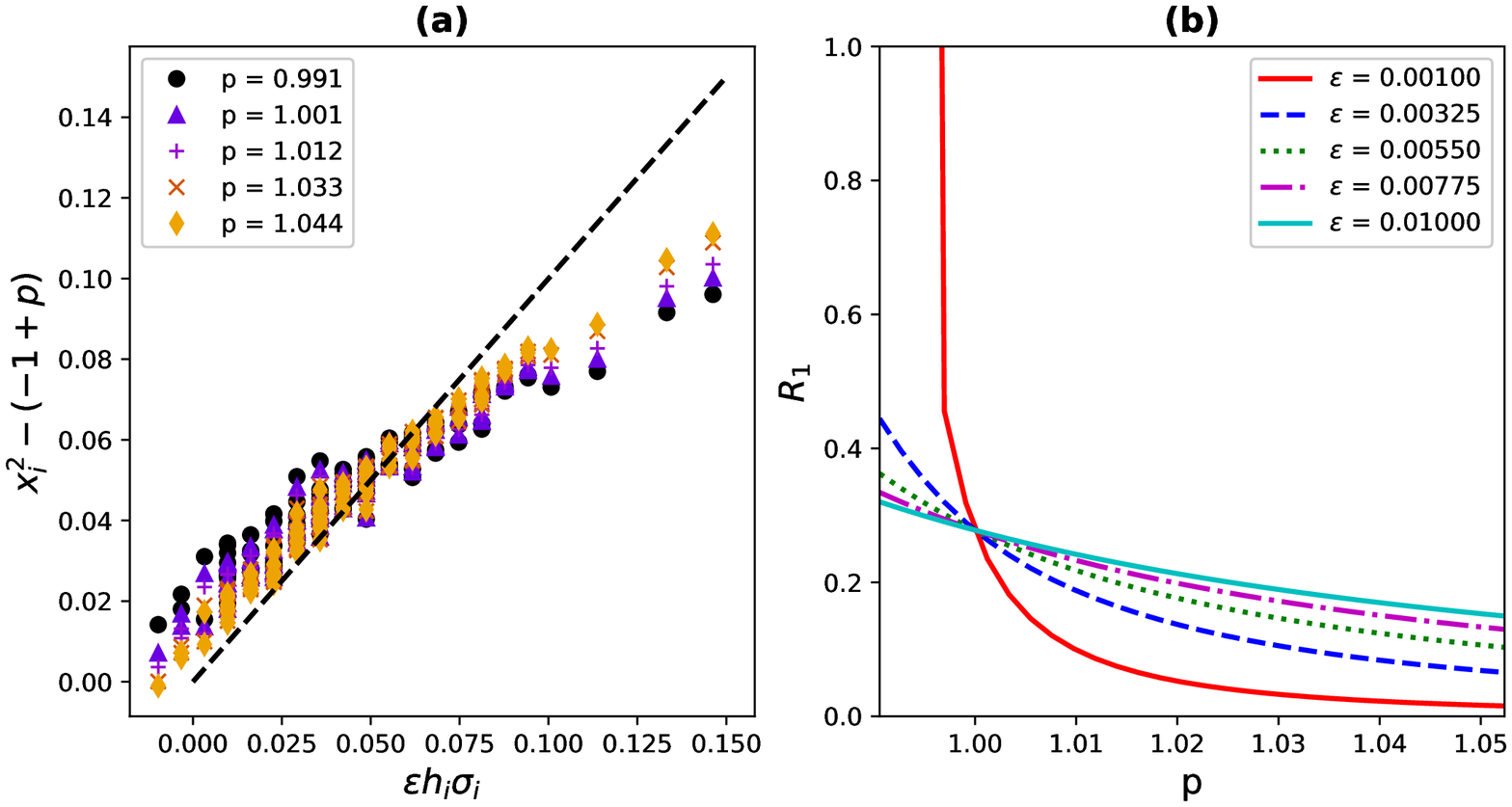}}
\caption{\label{fig:s1} Accuracy of the Taylor approximation of $x_i^2$ with $x_i^2 = (x_i^{(0)})^2 + 2 x_i^{(0)} x_i^{(1)} \epsilon + \mathcal{O}(\epsilon^2)$ when $e_i(0)=1$, $\forall i$. (a) First order term $x_i^2 - (x_i^{(0)})^2 =x_i^2 - (-1+p)$ vs.  $2 x_i^{(0)} x_i^{(1)} = \epsilon h_i \sigma_i$ obtained using numerical simulations of a Sherrington-Kirkpatrick spin glass problem of size $N=100$. Each marker corresponds to an index $i$ with $i \in \{1,\ldots,N\}$. (b) Normalized averaged remainder of the Taylor approximation $R_1$ with $R_1 = \frac{1}{N} \sum_i \frac{|x_i^2 - (-1+p) - \epsilon h_i \sigma_i|}{x_i^2}$ vs. the parameter value $p$. The remainder of the Taylor approximation is smoothly varying at the proximity of $p=1$ for a finite value of $\epsilon$.}
\end{figure*}


\subsection{Numerical simulation scheme and parameter values}

In the following, the scheme used for numerical simulations is detailed for the sake of reproducibility. The ODEs describing the time-evolution of the system are approximated using a Euler approximation given as follows (see eqs. (1) and (2)):

\begin{align}
x_i(t + \Delta t) &= x_i(t) + \Delta x_i(t) \Delta t, \forall i, \nonumber \tag{s10}\\
e_i(t + \Delta t) &= 
\begin{cases}
e_i(t) + \Delta e_i(t) \Delta t, \forall i, \text{ if } (<e_i(t) + \Delta e_i(t) \Delta t>) < \Gamma,\\
\frac{e_i(t) + \Delta e_i(t) \Delta t}{<e_i(t) + \Delta e_i(t) \Delta t>} \text{ otherwise. }
\end{cases} \nonumber \tag{s11}
\end{align}

\noindent  with $<X_i> = \frac{1}{N}\sum_i X_i$ and $\Gamma$ is the maximum mean error. Moreover, $\Delta x_i(t)$ and $\Delta e_i(t)$ are given as follows:

\begin{align}
\Delta x_i(t) &= (-1+p(t)) x_i(t) - x_i(t)^3 + \epsilon e_i(t) \sum_j \omega_{ij} x_j(t), \nonumber \tag{s12}\\
\Delta e_i(t) &= -\beta (x_i(t)^2 - a(t)) e_i(t), \nonumber \tag{s13}
\end{align}

If the configuration $\boldsymbol{\sigma}(t)$, defined as $\sigma_i(t) = \frac{x_i(t)}{|x_i(t)|}$, $\forall i$, is not a stable local mimimum, i.e., there exists an index $j$ such that $\sigma_j(t) h_j(t) < 0$ with $\boldsymbol{h}(t) = \Omega \boldsymbol{\sigma}(t)$, then the target amplitude $a(t)$ and linear gain $p(t)$ are given as follows (see eq. (9)):

\begin{align}
a(t) &= \alpha + \epsilon <e_i(t_c) h_i(t) \sigma_i(t)>, \nonumber \tag{s14}\\
p(t) &= \pi, \nonumber \tag{s15}
\end{align}

\noindent where $\pi$ is the linear gain baseline with $\pi<1$; $\alpha$, the target amplitude baseline; and $t_c$, with $t_c<t$, is the time of the last change of configuration $\boldsymbol{\sigma}$, i.e., the time at which one of the $x_i(t)$ has changed its sign.

If the configuration $\boldsymbol{\sigma}(t)$ is a local minimum, i.e., $\sigma_i(t) h_i(t) > 0$, $\forall i$, and $\theta(t) > \eta F(\mu_0(\boldsymbol{\sigma}(t))) $ (with $\mu_0(\boldsymbol{\sigma}(t)) < 2$, $\theta(t)$ given as $\theta(t) = \frac{a(t)}{1-p(t)}$, and $0 <\eta<1$) then $a(t)$ and $p(t)$ are given as follows (see eq. (7)):

\begin{align}
a(t) &=  \eta F(\mu_0(\boldsymbol{\sigma}(t))) (1-p(t)), \nonumber \tag{s16}\\
p(t) &= \pi + a(t)-a(t-1), \nonumber \tag{s17}
\end{align}

\noindent with $F(X) = -\frac{X}{X-2}$. Note that $\mu_0(\boldsymbol{\sigma}(t))$ is the dominant eigenvalue of the matrix $\tilde{\Omega} = D[(\boldsymbol{\sigma}(t) \cdot \boldsymbol{h}(t))^{-1}] \Omega - I$.

Lastly, the parameters of the system are described as given in Table 1 and the proposed scheme is simulated using Python on a Xeon E5-2680 2.70 GHz when solving the Sherrington-Kirkpatrick spin glass problems.

\begin{table}
  \begin{tabular}{ | l | c | r |}
    \hline
    Symbol & meaning & value \\ \hline \hline
		$\Delta t$ & Euler time-step & $0.05$  \\ 
    $\epsilon$ & coupling strength & $0.07$  \\ 
		$\alpha$ & target amplitude baseline & $0.2$  \\
		$\pi$ & linear gain baseline & $0.2$  \\
		$\beta$ & relative time-scale of error variables time-evolution & $0.05$  \\
		$\eta$ & destabilization strength from local minima & $0.5$  \\
		$\Gamma$ & maximal mean error & $5.0$  \\
		\hline
  \end{tabular}
	\caption{\label{tab0} Parameters description in the case of the Sherrington-Kirkpatrick spin glass problems.}
\end{table}

The time-evolution of the system is given as shown in the Supplementary Figure \ref{fig:s2}  when the parameter values from Table \ref{tab0} are used.

\begin{figure*}[htp]
\centerline{\includegraphics[width=1.0\textwidth]{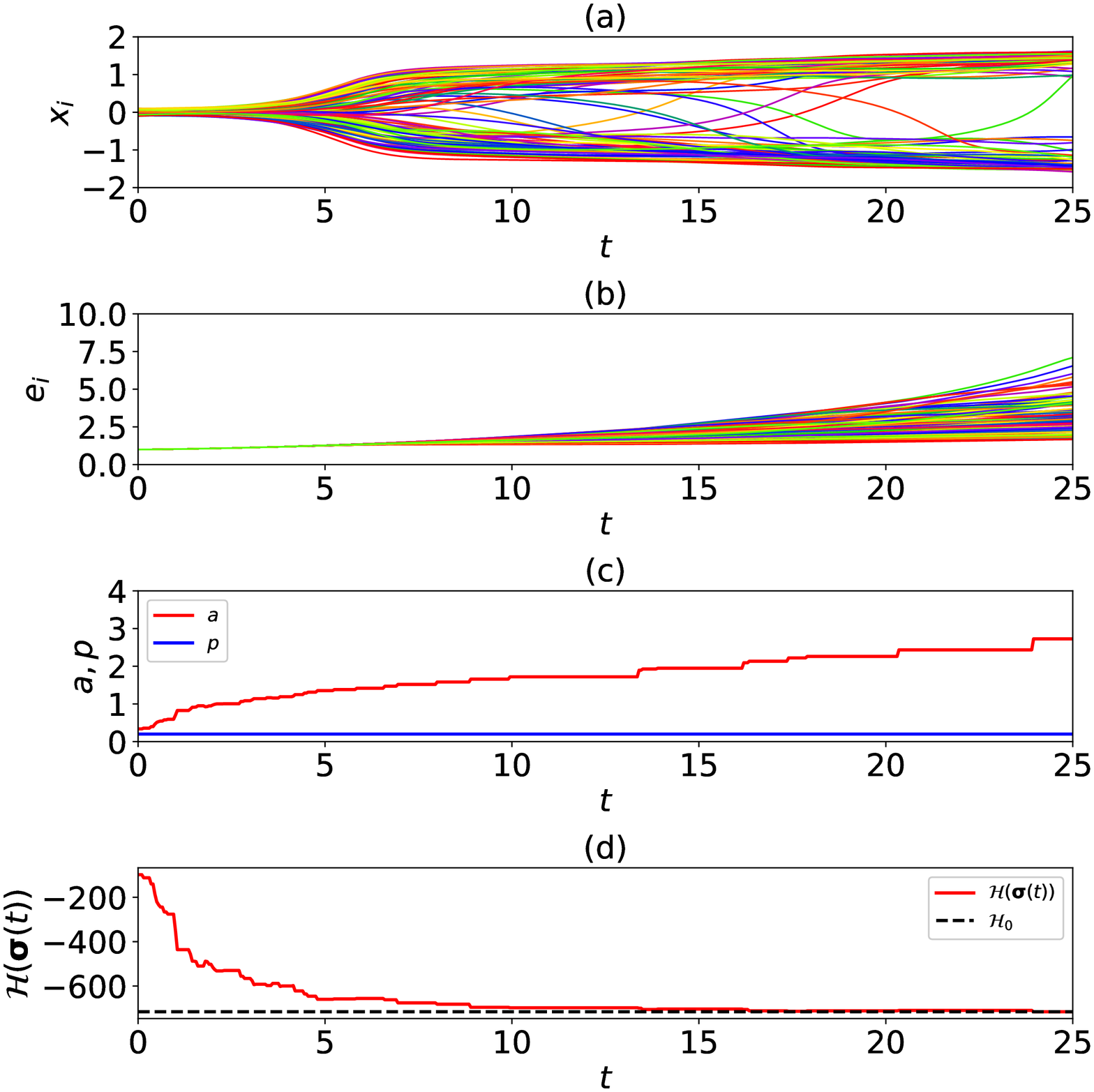}}
\caption{\label{fig:s2} Time-evolution of the states $x_i$, error variables $e_i$, target amplitude $a$, linear gain $p$, and Ising Hamiltonian $\mathcal{H}$ (cost function) are shown in (a), (b), (c), and (d), respectively, when solving a Sherrington-Kirkpatrick spin glass problem of size $N=100$. In (a) and (b), the different colors correspond different indices of $i$ with $i \in \{1,\ldots,N\}.$ At $t=25$, a ground-state configuration with energy $\mathcal{H}=\mathcal{H}_0$ has been found.}
\end{figure*}

\clearpage

\subsection{Benchmark results on the G-set}

In the case of the benchmark against BLS on the G-set, the proposed scheme is adapted as described in this section. First, the ODEs describing the system are simulated using a Euler approximation with time-step $\Delta t = 0.05$ (see eqs. [s10] and [s11]). Contrarily to the case of the Sherrington-Kirkpatrick problems, the error variables are not rescaled, i.e., $\Gamma = + \infty$. Note that the objective function in the case of the MAX-CUT problems is given as follows:

\begin{align}
\mathcal{C} = -\frac{1}{2}(\mathcal{H}+\frac{1}{2} \sum_{ij} \omega_{ij}), \nonumber \tag{s17}
\end{align}

\noindent where $\mathcal{C}$ is the value of the cut and $\mathcal{H}$ is the Ising Hamiltonian given as follows:

\begin{align}
\mathcal{H} = -\frac{1}{2} \sum_{ij} \omega_{ij} \sigma_i \sigma_j. \nonumber \tag{s18}
\end{align}

For MAX-CUT, the modulation of $\theta$ is achieved by modulating both the target amplitude $a$ and the pump rate $p$. First, the target amplitude is modulated as follows:

\begin{align}
a(t) = \alpha - \rho \phi(\delta \Delta \mathcal{C}(t)), \nonumber \tag{s19}
\end{align}

\noindent where $\Delta \mathcal{C}(t) = \mathcal{C}_{\text{opt}} - \mathcal{C}(t)$ with $\mathcal{C}_{\text{opt}} = -\frac{1}{2}(\mathcal{H}_{\text{opt}}+\frac{1}{2} \sum_{ij} \omega_{ij})$. Moreover, $\mathcal{C}(t)$ is the cut of the currently visited configuration and $\mathcal{H}_{\text{opt}}$ the target energy. In practice, we set $\mathcal{H}_{\text{opt}}$ to the lowest energy visited, i.e., $\mathcal{H}_{\text{opt}}(t) = \text{min}_{t' \leq t} \mathcal{H}(t')$. Lastly, the function $\phi$ is the tangent hyperbolic function. Next, the pump rate is modulated as follows:

\begin{align} \label{eq:pumpEnergy}
p(t) = \pi + \rho \phi (\delta \Delta \mathcal{C}(t)). \nonumber \tag{s20}
\end{align}

In this case, the parameter $\beta$ is time-dependent. It is linearly increased with a rate equal to $\gamma$ during the simulation, and reset to zero if the energy does not decrease during a duration $\tau$ with $\tau > 0$. The dynamics of $\beta(t)$ is given as follows ($\beta(0) = 0$):

\begin{align}
\partial_t \beta = \gamma, \nonumber \tag{s21}
\end{align}

\noindent when $t - t_c < \tau$, with $t_c$ the last time the best known cut $\mathcal{C}_{\text{opt}}$ increased or $\beta$ was reset. Otherwise, $\beta$ is reset to $0$ if $t - t_c \geq \tau$ and $t_c$ is set to $t$. The parameters used for the numerical simulations of the proposed scheme are summarized in Tab. \ref{tab0b}. Moreover, the function $g$ is given as $g(\nu_1,\nu_2,\nu_3,\nu_4) = -6.3674 - 0.2579 \nu_1  -1.0548 \nu_2 -4.2597 \nu_3 + 6.1727 \nu_4$ where $\nu_1$, $\nu_2$, $\nu_3$, and $\nu_4$ are the largest, second, third, and fourth largest eigenvalues, respectively, of the weight matrix $\Omega=\{\omega_{ij}\}_{ij}$.

\begin{table}
  \begin{tabular}{ | l | c | r |}
    \hline
    Symbol & meaning & value \\ \hline \hline
    $\epsilon$ & coupling strength & $3/<\sum_i |\omega_{ij}| >$  \\ 
		$\alpha$ & target amplitude baseline & 1  \\
		$\pi$ & linear gain baseline & $1 - \epsilon g(\nu_1,\nu_2,\nu_3,\nu_4)$  \\
		$\rho$ & amplitude and gain variation & 1  \\
		$\delta$ & sensitivity to energy variations & 7  \\
		$\gamma$ & rate of increase of $\beta$ & 0.065 / N  \\
		$\tau$ & max. time w/o energy change & 7 N \\
		\hline
  \end{tabular}
	\caption{\label{tab0b} Parameters description in the case of MAX-CUT on the G-set.}
\end{table}

Results of the benchmark on the G-set are shown in Tables \ref{tab1}, \ref{tab2}, and \ref{tab3}.

\begin{table*}
  \begin{tabular}{ | c | c |  c |  c |  c | c | c | c |}
	\hline
id & N & $C^{\text{BLS}}$ & $C^{*}$ & $C^{*}-C^{\text{BLS}}$ & $<t^{\text{BLS}}>$ (s) & $<t^{\text{CPU}}>$ (s) & $<t^{\text{CIM}}>$ (s) \\ \hline
1 & 800 & 11624(9) & 11624(20) & 0 & 13.00 & 12.69  & 0.03 \\
2 & 800 & 11620(11) & 11620(20) & 0 & 41.00 & 49.07  & 0.13 \\ 
3 & 800 & 11622(19) & 11622(20) & 0 & 83.00 & 24.16  & 0.06 \\ 
4 & 800 & 11646(16) & 11646(20) & 0 & 214.00 & 38.26  & 0.10  \\
5 & 800 & 11631(20) & 11631(19) & 0 & 14.00 & 25.80  & 0.06  \\
6 & 800 & 2178(20) & 2178(20) & 0 & 18.00 & 14.74  & 0.04  \\
7 & 800 & 2006(12) & 2006(20) & 0 & 317.00 & 32.19  & 0.08  \\
8 & 800 & 2005(18) & 2005(18) & 0 & 195.00 & 75.20  & 0.18  \\
9 & 800 & 2054(14) & 2054(20) & 0 & 97.00 & 34.26  & 0.09  \\
10 & 800 & 2000(13) & 2000(19) & 0 & 79.00 & 134.81  & 0.30  \\
11 & 800 & 564(20) & 564(20) & 0 & 1.00 & 21.19  & 0.07  \\
12 & 800 & 556(20) & 556(20) & 0 & 2.00 & 12.40  & 0.04  \\
13 & 800 & 582(20) & 582(20) & 0 & 2.00 & 31.10  & 0.10  \\
14 & 800 & 3064(6) & 3064(9) & 0 & 119.00 & 107.89  & 0.34  \\
15 & 800 & 3050(20) & 3050(6) & 0 & 43.00 & 37.72  & 0.10  \\
16 & 800 & 3052(8) & 3052(20) & 0 & 70.00 & 89.18  & 0.27  \\
17 & 800 & 3047(16) & 3047(17) & 0 & 96.00 & 131.32  & 0.39  \\
18 & 800 & 992(14) & 992(7) & 0 & 106.00 & 72.14  & 0.19  \\
19 & 800 & 906(6) & 906(15) & 0 & 20.00 & 101.04  & 0.29  \\
20 & 800 & 941(20) & 941(19) & 0 & 9.00 & 37.86  & 0.10  \\
21 & 800 & 931(14) & 931(9) & 0 & 42.00 & 109.44  & 0.32  \\
  		\hline
  \end{tabular}
\caption{\label{tab1} Performance of proposed method in finding ground-states, i.e., maximum cuts, of graphs in the G-set. $id$, $C^{\text{BLS}}$, and $C^{*}$ are the name of instances, maximum cuts found by BLS\cite{Benlic2013s} and the proposed method, respectively, after 20 runs. The number of runs that found the optimal cut are indicated in parenthesis. Moreover, $<t^{\text{BLS}}>$, $<t^{\text{CPU}}>$, and $<t^{\text{CIM}}>$ are the averaged time-to-solution using BLS written C++ and running on a Xeon E5440 2.83 GHz\cite{Benlic2013s}, the proposed scheme simulated using Matlab on a Xeon X5690 3.47 GHz, and the expected time-to-solution for an implementation of the proposed scheme using the coherent Ising machine\cite{McMahon2016s} (see footnote [45]), respectively. Part I.}
\end{table*}

\begin{table*}
  \begin{tabular}{ | c | c |  c |  c |  c | c | c | c |}
	\hline
id & N & $C^{\text{BLS}}$ & $C^{*}$ & $C^{*}-C^{\text{BLS}}$ & $<t^{\text{BLS}}>$ (s) & $<t^{\text{CPU}}>$ (s) & $<t^{\text{CIM}}>$ (s) \\ \hline
22 & 2000 & 13359(1) & 13359(9) & 0 & 560.00 & 229.53  & 0.43  \\
23 & 2000 & 13344(10) & 13342(11) & -2 & 278.00 & 383.30  & 0.71 \\ 
24 & 2000 & 13337(5) & 13337(17) & 0 & 311.00 & 493.25  & 0.90  \\
25 & 2000 & 13340(1) & 13340(10) & 0 & 148.00 & 414.78  & 0.78  \\
26 & 2000 & 13328(3) & 13328(8) & 0 & 429.00 & 534.15  & 0.98  \\
27 & 2000 & 3341(10) & 3341(20) & 0 & 449.00 & 228.86  & 0.43  \\
28 & 2000 & 3298(8) & 3298(16) & 0 & 432.00 & 434.56  & 0.78  \\
29 & 2000 & 3405(1) & 3405(19) & 0 & 17.00 & 526.89  & 0.90  \\
30 & 2000 & 3412(10) & 3413(1) & 1 & 283.00 & 164.02  & 0.29  \\
31 & 2000 & 3309(8) & 3310(2) & 1 & 285.00 & 483.15  & 0.87  \\
32 & 2000 & 1410(13) & 1408(7) & -2 & 336.00 & 422.86  & 0.97  \\
33 & 2000 & 1382(1) & 1380(3) & -2 & 402.00 & 320.80  & 0.75  \\
34 & 2000 & 1384(20) & 1384(10) & 0 & 170.00 & 410.30  & 0.96  \\
35 & 2000 & 7684(2) & 7681(1) & -3 & 442.00 & 655.11  & 1.47  \\
36 & 2000 & 7678(1) & 7676(8) & -2 & 604.00 & 251.63  & 0.48  \\
37 & 2000 & 7689(1) & 7690(1) & 1 & 444.00 & 285.74  & 0.54  \\
38 & 2000 & 7687(4) & 7688(1) & 1 & 461.00 & 199.44  & 0.38  \\
39 & 2000 & 2408(13) & 2408(5) & 0 & 251.00 & 438.54  & 0.84  \\
40 & 2000 & 2400(1) & 2400(1) & 0 & 431.00 & 528.28  & 1.10  \\
41 & 2000 & 2405(17) & 2405(3) & 0 & 73.00 & 132.46  & 0.24  \\
42 & 2000 & 2481(2) & 2478(1) & -3 & 183.00 & 199.26  & 0.35  \\
 		\hline
  \end{tabular}
\caption{\label{tab2} Same as Tab. \ref{tab1}, part II.}
\end{table*}

\begin{table*}
  \begin{tabular}{ | c | c |  c |  c |  c | c | c | c |}
	\hline
id & N & $C^{\text{BLS}}$ & $C^{*}$ & $C^{*}-C^{\text{BLS}}$ & $<t^{\text{BLS}}>$ (s) & $<t^{\text{CPU}}>$ (s) & $<t^{\text{CIM}}>$ (s) \\ \hline
43 & 1000 & 6660(18) & 6660(20) & 0 & 26.00 & 30.01  & 0.08  \\
44 & 1000 & 6650(14) & 6650(20) & 0 & 43.00 & 46.35  & 0.12  \\
45 & 1000 & 6654(15) & 6654(20) & 0 & 104.00 & 104.78  & 0.27  \\
46 & 1000 & 6649(13) & 6649(20) & 0 & 67.00 & 110.58  & 0.30  \\
47 & 1000 & 6657(11) & 6657(20) & 0 & 102.00 & 102.36  & 0.28  \\
48 & 3000 & 6000(20) & 6000(19) & 0 & 0.00 & 4.76  & 0.01  \\
49 & 3000 & 6000(20) & 6000(12) & 0 & 0.00 & 4.99  & 0.01  \\
50 & 3000 & 5880(19) & 5880(6) & 0 & 169.00 & 18.40  & 0.04  \\
51 & 1000 & 3848(17) & 3848(8) & 0 & 81.00 & 181.31  & 0.49  \\
52 & 1000 & 3851(19) & 3851(3) & 0 & 78.00 & 199.90  & 0.62  \\
53 & 1000 & 3850(13) & 3850(6) & 0 & 117.00 & 187.95  & 0.51  \\
54 & 1000 & 3852(10) & 3851(19) & -1 & 131.00 & 201.95  & 0.60  \\
 		\hline
  \end{tabular}
\caption{\label{tab3} Same as Tab. \ref{tab1}, part III.}
\end{table*}

\clearpage

For instances 30, 31, 37, and 38, the proposed scheme finds solutions of better quality than previously known from \cite{Benlic2013s}. The solutions found are given as follows.

$\bullet$ Solution of instance 30 with cut 3413:

-1 -1 1 1 -1 1 -1 -1 1 1 -1 1 1 1 -1 -1 -1 1 -1 1 1 1 -1 1 -1 1 1 -1 1 -1 -1 1 -1 1 -1 1 -1 -1 1 1 1 -1 -1 1 1 -1 1 -1 1 1 -1 1 -1 1 -1 -1 1 1 1 -1 1 -1 -1 1 -1 -1 1 -1 1 -1 -1 1 -1 -1 -1 -1 1 1 -1 -1 1 1 -1 1 1 1 -1 1 1 -1 1 1 -1 1 -1 1 1 1 -1 -1 -1 -1 -1 -1 1 -1 1 -1 1 1 1 1 -1 -1 1 1 1 -1 -1 -1 -1 1 1 -1 1 1 1 -1 -1 -1 -1 -1 -1 1 -1 1 1 1 1 1 1 1 -1 -1 1 1 1 1 -1 1 1 -1 1 -1 -1 1 1 1 1 -1 1 -1 -1 -1 -1 1 -1 -1 1 1 -1 1 1 1 1 1 1 -1 -1 1 1 1 -1 -1 1 1 1 -1 -1 -1 1 1 -1 -1 1 -1 1 1 1 -1 -1 1 -1 -1 1 1 1 -1 1 1 -1 -1 -1 1 1 -1 1 1 -1 1 -1 1 -1 -1 1 1 -1 1 1 -1 -1 -1 -1 1 -1 1 -1 1 -1 -1 -1 -1 -1 -1 1 -1 -1 1 -1 1 1 1 -1 1 1 -1 1 1 -1 1 -1 -1 -1 1 1 -1 -1 -1 1 1 1 1 -1 1 1 -1 -1 -1 -1 -1 -1 -1 1 -1 1 1 1 1 1 -1 -1 -1 -1 -1 1 1 1 1 1 1 1 -1 -1 1 -1 1 -1 1 -1 1 1 1 -1 -1 -1 -1 1 1 -1 1 -1 -1 1 1 1 -1 1 -1 1 -1 -1 -1 -1 1 -1 -1 -1 1 1 -1 -1 1 1 -1 -1 -1 1 -1 -1 1 1 -1 1 -1 1 -1 1 1 -1 1 -1 1 -1 1 -1 1 1 -1 1 -1 1 -1 1 1 -1 -1 1 1 1 -1 1 1 -1 1 -1 -1 -1 -1 1 1 1 1 1 1 1 -1 1 1 -1 1 1 1 1 -1 1 1 1 1 -1 -1 1 -1 -1 -1 -1 -1 -1 1 -1 -1 1 1 -1 1 -1 -1 1 1 1 -1 -1 -1 1 -1 1 -1 1 1 -1 -1 1 1 1 -1 -1 -1 1 1 1 -1 1 -1 1 1 -1 -1 1 -1 1 -1 -1 -1 1 -1 -1 -1 1 -1 -1 1 1 -1 -1 -1 1 -1 1 1 1 1 -1 1 -1 -1 -1 1 1 -1 1 -1 1 1 1 1 1 1 1 -1 -1 1 1 1 1 1 -1 1 1 -1 -1 1 -1 -1 -1 1 1 1 -1 -1 1 1 1 -1 -1 1 -1 -1 -1 1 -1 1 -1 -1 1 -1 1 -1 -1 1 -1 -1 1 -1 1 1 1 -1 1 -1 -1 1 -1 1 1 -1 -1 1 1 -1 -1 1 -1 -1 1 -1 -1 -1 1 1 1 1 1 1 1 -1 1 1 -1 -1 -1 1 1 1 1 -1 1 1 1 -1 -1 1 1 1 1 -1 1 1 -1 1 1 -1 -1 1 -1 1 -1 -1 1 1 -1 -1 1 -1 1 -1 1 -1 -1 1 -1 1 1 1 1 1 1 1 1 1 1 1 1 -1 1 -1 -1 -1 -1 -1 1 -1 -1 1 1 1 -1 1 1 1 1 1 -1 1 1 1 -1 -1 -1 -1 -1 -1 1 -1 -1 1 -1 1 -1 1 1 1 1 1 1 -1 1 -1 -1 1 1 1 -1 1 1 -1 -1 -1 1 1 -1 1 1 -1 -1 1 -1 1 -1 1 -1 1 -1 1 -1 1 1 -1 1 1 -1 -1 -1 1 1 1 -1 1 -1 -1 1 -1 -1 1 1 -1 1 -1 1 -1 1 -1 -1 1 1 1 1 1 1 -1 -1 -1 -1 -1 -1 1 -1 -1 1 -1 1 -1 1 -1 1 -1 -1 -1 1 -1 -1 -1 1 -1 1 -1 -1 -1 -1 1 -1 -1 -1 -1 1 -1 1 1 1 1 1 1 -1 -1 1 1 -1 1 -1 1 -1 -1 -1 1 1 1 -1 1 -1 -1 1 -1 1 1 -1 -1 -1 1 1 -1 -1 -1 1 1 -1 1 -1 -1 -1 -1 1 -1 -1 -1 1 1 -1 1 1 -1 -1 -1 -1 -1 1 1 -1 1 1 1 -1 1 1 -1 -1 1 -1 1 1 1 1 1 -1 1 -1 -1 -1 1 -1 -1 1 -1 1 1 1 1 -1 -1 -1 -1 1 -1 1 -1 1 1 -1 1 1 -1 -1 1 1 1 -1 1 1 1 1 1 -1 1 1 -1 -1 1 -1 -1 1 -1 -1 -1 -1 1 -1 1 1 -1 -1 -1 1 1 -1 -1 1 1 1 1 1 -1 -1 1 -1 -1 -1 1 1 1 1 -1 -1 1 -1 -1 -1 -1 1 1 -1 1 -1 1 1 -1 1 -1 1 -1 -1 -1 -1 -1 -1 -1 -1 -1 1 1 -1 -1 -1 1 -1 -1 1 -1 1 -1 -1 -1 -1 -1 -1 -1 1 1 1 -1 -1 -1 1 1 -1 1 -1 1 -1 -1 -1 -1 1 -1 1 1 1 -1 -1 -1 -1 1 -1 -1 -1 1 -1 1 1 1 -1 1 -1 1 -1 -1 -1 1 1 1 -1 -1 1 1 1 -1 1 1 1 1 -1 1 -1 1 1 1 -1 -1 1 -1 -1 -1 1 -1 1 -1 -1 -1 -1 1 1 -1 1 1 1 1 1 1 1 1 1 1 -1 -1 1 1 1 -1 -1 -1 1 -1 1 -1 -1 1 1 -1 1 -1 1 -1 -1 -1 -1 -1 -1 1 1 1 -1 -1 1 -1 -1 1 1 1 -1 -1 -1 -1 1 1 1 1 1 1 1 1 -1 -1 -1 -1 1 -1 -1 1 1 1 -1 1 -1 -1 -1 -1 -1 1 1 1 -1 -1 1 -1 -1 -1 1 1 1 1 1 -1 1 -1 -1 1 -1 1 1 1 1 -1 1 -1 -1 1 1 -1 -1 1 1 1 1 -1 1 -1 -1 -1 -1 -1 -1 1 1 -1 1 -1 1 -1 1 1 1 -1 -1 -1 1 -1 -1 -1 -1 -1 -1 -1 -1 1 1 -1 -1 1 1 1 1 -1 1 1 1 1 -1 1 1 1 1 1 1 1 -1 1 -1 1 -1 1 1 -1 -1 -1 -1 1 -1 -1 -1 1 1 -1 -1 -1 1 -1 -1 1 1 1 -1 -1 -1 1 -1 1 1 1 -1 1 -1 1 1 -1 -1 -1 1 1 -1 1 1 1 1 1 -1 1 1 1 1 1 1 1 -1 1 1 1 1 -1 -1 -1 1 -1 1 -1 1 1 1 -1 -1 1 -1 -1 -1 1 1 1 -1 1 1 1 1 -1 1 1 -1 -1 -1 -1 1 -1 1 -1 1 1 1 1 -1 -1 -1 -1 -1 1 -1 1 1 -1 -1 -1 1 -1 -1 1 1 1 -1 -1 -1 1 1 -1 1 -1 1 1 1 -1 1 -1 1 -1 1 1 -1 -1 1 -1 -1 -1 -1 -1 1 -1 1 1 -1 -1 -1 1 1 1 1 -1 -1 1 1 1 1 -1 1 -1 1 1 1 -1 1 1 -1 -1 -1 -1 1 1 1 -1 1 -1 1 1 -1 -1 -1 1 -1 1 -1 1 -1 1 -1 1 1 -1 1 -1 -1 1 1 1 1 1 1 1 -1 -1 1 1 -1 -1 1 -1 1 1 -1 -1 -1 1 -1 1 -1 1 -1 -1 -1 1 1 -1 -1 -1 1 -1 1 1 -1 -1 1 1 1 1 1 -1 1 -1 -1 -1 -1 -1 -1 -1 1 1 1 -1 1 1 1 1 -1 -1 -1 -1 1 1 -1 -1 -1 -1 1 -1 -1 -1 1 -1 -1 -1 -1 1 1 1 1 1 1 -1 1 1 -1 1 -1 -1 1 1 1 -1 -1 -1 1 -1 1 -1 1 1 -1 1 -1 -1 1 -1 1 -1 1 1 1 -1 -1 1 1 1 -1 1 1 1 1 -1 -1 1 -1 -1 1 1 -1 1 1 1 1 -1 1 -1 -1 -1 1 1 -1 1 -1 1 1 -1 1 1 1 1 1 1 -1 -1 1 -1 -1 -1 1 -1 -1 1 1 -1 1 1 -1 -1 1 1 1 1 -1 -1 1 1 -1 1 1 1 1 -1 -1 1 -1 -1 1 1 -1 1 1 1 -1 -1 -1 1 1 -1 1 1 1 1 1 -1 -1 -1 -1 1 -1 1 -1 -1 -1 -1 1 -1 -1 -1 1 1 1 -1 1 -1 1 1 1 -1 1 1 -1 1 -1 -1 1 -1 -1 1 -1 1 -1 1 1 1 -1 -1 1 1 -1 1 1 1 -1 -1 1 1 -1 1 1 -1 1 1 1 -1 -1 -1 1 -1 1 -1 1 1 1 -1 -1 1 -1 -1 -1 -1 1 1 -1 -1 1 -1 1 -1 -1 1 1 -1 -1 1 -1 -1 -1 -1 -1 -1 -1 -1 -1 -1 1 1 1 1 -1 -1 -1 1 1 1 -1 -1 -1 -1 1 1 1 -1 -1 1 -1 1 1 -1 1 1 -1 -1 -1 -1 -1 1 -1 1 -1 -1 -1 1 1 1 1 -1 -1 -1 -1 -1 1 -1 1 -1 -1 -1 -1 1 -1 1 1 1 1 1 -1 1 1 -1 1 1 -1 1 1 1 -1 1 1 1 1 1 1 1 1 1 1 -1 1 1 -1 1 1 1 -1 -1 1 -1 1 -1 -1 1 1 1 1 1 1 1 -1 1 1 -1 1 -1 1 1 -1 1 -1 -1 1 1 1 -1 -1 -1 -1 -1 1 -1 1 1 -1 -1 1 -1 1 1 1 1 -1 -1 1 1 1 1 1 1 -1 1 1 -1 1 -1 -1 1 -1 1 1 1 -1 1 1 -1 -1 -1 1 1 -1 -1 -1 -1 -1 -1 -1 -1 -1 -1 -1 -1 -1 1 1 1 1 -1 -1 1 1 -1 1 -1 1 -1 -1 -1 1 -1 -1 -1 -1 -1 1 -1 1 1 1 -1 1 -1 1 1 1 -1 -1 -1 1 -1 1 1 -1 1 1 -1 1 -1 1 -1 -1 -1 -1 1 1 -1 1 1 -1 -1 1 -1 -1 -1 1 1 -1 -1 -1 1 -1 -1 -1 -1 -1 -1 -1 -1 -1 -1 1 1 -1 1 -1 -1 1 -1 -1 1 1 -1 -1 -1 1 1 -1 1 -1 -1 -1 -1 1 -1 -1 1 1 -1

$\bullet$ Solution of instance 31 with cut 3310:

-1 1 1 -1 1 -1 1 1 1 -1 1 -1 1 -1 1 1 1 -1 -1 -1 -1 1 1 -1 1 1 1 1 1 -1 1 1 -1 -1 -1 1 1 1 1 -1 1 1 1 1 -1 1 -1 1 1 1 -1 1 -1 1 -1 -1 1 -1 -1 -1 -1 1 -1 -1 1 1 -1 -1 -1 1 -1 -1 -1 -1 -1 1 -1 1 -1 1 -1 -1 -1 1 -1 -1 1 1 -1 1 1 1 1 1 -1 1 1 1 1 1 1 -1 1 -1 -1 -1 1 1 -1 -1 -1 -1 -1 -1 -1 -1 -1 -1 -1 1 1 -1 -1 -1 -1 1 -1 1 -1 1 1 1 1 -1 1 -1 -1 1 1 -1 -1 -1 -1 1 -1 1 1 -1 1 1 -1 -1 1 -1 -1 -1 -1 -1 1 1 1 -1 -1 -1 1 1 1 -1 -1 -1 1 1 1 1 -1 1 -1 -1 -1 1 1 1 -1 -1 1 -1 -1 -1 1 -1 -1 -1 -1 1 -1 -1 -1 -1 1 1 -1 -1 -1 1 1 -1 -1 1 -1 1 1 -1 -1 1 1 -1 1 -1 1 -1 -1 1 1 -1 -1 -1 1 -1 1 1 -1 1 1 1 1 -1 -1 -1 -1 -1 -1 1 -1 -1 1 -1 -1 1 -1 -1 -1 -1 -1 1 1 1 1 1 1 -1 1 1 1 -1 1 1 -1 1 -1 -1 1 -1 1 -1 -1 1 1 -1 1 -1 -1 1 -1 -1 1 -1 1 1 1 -1 -1 -1 -1 1 1 1 -1 -1 1 1 -1 1 -1 1 1 1 1 1 1 -1 -1 -1 1 -1 -1 -1 -1 1 1 1 -1 -1 -1 1 -1 1 -1 1 1 1 1 1 1 1 1 1 -1 -1 -1 1 -1 1 -1 -1 1 -1 -1 -1 1 -1 -1 1 -1 -1 -1 -1 1 -1 1 1 -1 -1 1 -1 1 1 -1 -1 -1 -1 -1 1 1 -1 1 -1 -1 1 -1 -1 -1 -1 1 1 -1 -1 1 1 -1 -1 1 -1 -1 -1 -1 1 -1 -1 1 1 1 1 -1 -1 -1 1 1 -1 -1 -1 1 -1 -1 1 -1 -1 1 1 1 -1 1 -1 -1 -1 -1 -1 -1 1 -1 1 -1 -1 -1 1 -1 1 -1 -1 1 1 -1 1 1 -1 1 1 -1 -1 -1 -1 1 -1 -1 -1 1 1 -1 -1 1 -1 1 -1 1 -1 -1 -1 -1 -1 1 -1 -1 1 1 1 -1 -1 -1 -1 -1 -1 1 1 1 -1 1 -1 -1 -1 -1 -1 -1 1 -1 1 -1 1 -1 1 1 1 -1 1 -1 1 1 -1 1 1 -1 1 1 1 -1 1 -1 1 -1 1 1 1 1 1 -1 -1 -1 -1 1 1 -1 -1 1 1 1 -1 1 -1 1 1 1 -1 -1 1 1 -1 1 -1 -1 -1 -1 -1 -1 -1 -1 -1 1 -1 1 1 -1 1 -1 -1 1 1 1 -1 -1 1 -1 1 -1 1 -1 1 1 -1 -1 1 1 -1 1 -1 -1 1 -1 -1 1 -1 -1 -1 -1 -1 -1 -1 1 1 -1 -1 1 -1 -1 -1 -1 1 1 -1 1 -1 -1 -1 1 -1 1 1 -1 1 1 -1 1 -1 -1 -1 -1 -1 1 -1 1 -1 1 -1 1 -1 -1 1 -1 -1 1 -1 -1 -1 1 -1 1 -1 1 -1 -1 -1 -1 1 -1 1 1 -1 1 1 1 -1 -1 -1 -1 1 -1 -1 1 1 1 1 1 1 1 1 1 1 1 1 1 -1 -1 -1 1 1 -1 -1 -1 1 1 -1 1 1 1 -1 -1 -1 -1 1 1 1 -1 -1 -1 -1 1 -1 -1 -1 -1 1 1 -1 1 -1 1 -1 -1 -1 1 1 1 -1 1 -1 1 1 -1 1 -1 -1 -1 -1 -1 -1 -1 1 1 -1 1 1 -1 1 -1 1 -1 -1 -1 1 -1 -1 -1 1 1 -1 1 1 1 -1 -1 1 1 1 1 -1 1 -1 -1 -1 1 -1 -1 -1 1 -1 1 1 -1 1 -1 1 -1 -1 1 -1 1 -1 1 1 -1 1 -1 1 1 -1 1 -1 1 1 -1 -1 1 1 -1 -1 1 -1 -1 -1 -1 1 1 -1 -1 -1 1 1 1 1 -1 -1 -1 -1 1 1 -1 -1 -1 1 -1 -1 1 1 -1 1 -1 -1 1 1 1 -1 -1 -1 1 1 -1 1 1 -1 -1 -1 1 -1 -1 -1 1 1 -1 1 1 1 -1 1 1 -1 -1 -1 -1 -1 -1 -1 1 1 1 1 -1 1 1 -1 -1 1 -1 1 1 1 1 1 1 1 -1 1 -1 1 -1 1 1 -1 1 1 -1 -1 -1 -1 -1 -1 1 1 1 -1 1 -1 1 -1 -1 1 1 -1 1 -1 1 -1 1 1 1 1 1 1 -1 1 -1 1 -1 1 -1 1 -1 1 -1 1 -1 -1 1 -1 -1 -1 1 -1 -1 -1 -1 -1 1 1 1 1 -1 1 -1 1 1 -1 1 -1 1 -1 1 1 1 -1 -1 -1 1 1 1 1 1 1 -1 1 -1 -1 -1 -1 -1 -1 -1 -1 1 1 1 1 1 -1 -1 1 1 1 -1 1 -1 1 1 -1 1 1 -1 -1 -1 -1 1 1 -1 1 1 1 -1 -1 -1 -1 1 -1 -1 1 -1 1 1 -1 -1 1 -1 -1 -1 1 1 -1 -1 1 1 -1 -1 -1 1 -1 1 1 1 -1 -1 -1 -1 -1 1 -1 1 -1 1 1 1 1 1 -1 -1 1 1 1 -1 -1 -1 -1 -1 1 1 1 -1 -1 -1 1 -1 -1 1 -1 1 -1 1 1 -1 -1 -1 -1 1 -1 1 -1 -1 1 -1 1 -1 1 -1 -1 -1 -1 -1 1 -1 -1 1 1 -1 1 1 1 1 -1 -1 -1 1 -1 1 -1 -1 -1 1 -1 -1 -1 -1 -1 -1 -1 -1 1 -1 1 1 -1 1 -1 1 1 -1 -1 1 -1 1 -1 1 1 1 1 -1 1 1 1 -1 1 -1 1 -1 1 -1 1 -1 1 -1 1 -1 1 1 1 -1 1 -1 1 -1 -1 1 -1 1 -1 1 -1 1 1 -1 1 -1 1 -1 1 1 1 1 1 1 -1 1 1 1 -1 -1 -1 1 1 -1 -1 -1 -1 1 1 1 1 -1 1 1 1 1 1 1 -1 -1 -1 -1 1 -1 1 1 1 1 1 1 1 -1 -1 1 1 1 -1 1 -1 1 -1 1 -1 1 -1 -1 1 1 -1 1 -1 -1 1 -1 -1 -1 1 1 -1 -1 1 1 1 -1 -1 -1 -1 -1 1 -1 1 1 1 -1 -1 1 -1 -1 1 1 -1 -1 1 -1 1 1 1 1 -1 1 -1 -1 1 -1 1 -1 1 1 1 1 1 1 1 1 -1 1 1 1 -1 1 -1 1 -1 -1 1 -1 1 -1 -1 1 1 -1 1 1 -1 -1 -1 -1 1 1 1 -1 -1 -1 -1 -1 -1 1 -1 -1 -1 -1 1 1 -1 1 1 -1 1 -1 1 1 1 1 -1 -1 -1 -1 -1 -1 -1 -1 1 -1 -1 -1 -1 -1 1 1 -1 -1 -1 1 1 -1 -1 -1 -1 1 -1 1 -1 -1 1 1 -1 -1 1 -1 1 1 1 1 1 1 1 -1 -1 1 1 1 1 -1 1 1 -1 1 1 -1 1 -1 1 1 -1 1 1 1 -1 1 1 -1 -1 -1 1 -1 -1 -1 -1 -1 -1 1 -1 -1 -1 -1 1 1 1 -1 1 -1 1 -1 1 1 1 -1 1 1 1 1 1 -1 1 -1 -1 1 1 -1 -1 1 1 1 -1 1 1 1 1 1 -1 1 1 1 1 -1 1 -1 -1 1 -1 1 1 -1 -1 -1 -1 1 -1 1 -1 -1 1 -1 -1 -1 -1 -1 1 1 -1 -1 -1 -1 -1 1 1 1 1 1 1 1 1 1 -1 1 -1 -1 1 -1 1 -1 -1 -1 1 -1 -1 1 1 1 -1 -1 -1 -1 1 -1 -1 -1 -1 1 -1 -1 1 1 -1 -1 1 1 -1 -1 1 1 -1 -1 1 1 -1 -1 -1 -1 -1 -1 -1 1 1 -1 1 -1 -1 -1 -1 -1 -1 1 1 -1 -1 -1 1 -1 1 -1 1 1 1 1 1 -1 1 -1 1 -1 1 -1 -1 1 1 1 1 -1 -1 -1 1 -1 1 1 -1 1 -1 -1 1 -1 1 1 1 1 -1 -1 1 1 -1 -1 1 1 -1 1 -1 -1 1 1 1 -1 1 -1 -1 1 1 -1 -1 -1 1 -1 1 -1 1 -1 1 -1 1 -1 -1 1 -1 -1 1 1 -1 -1 -1 -1 1 -1 -1 -1 -1 -1 1 -1 1 1 -1 1 -1 1 1 -1 -1 -1 -1 1 -1 -1 1 1 -1 -1 1 -1 -1 -1 1 -1 1 1 1 -1 -1 -1 1 1 -1 -1 1 1 1 -1 -1 -1 1 -1 -1 1 1 1 1 1 -1 -1 -1 -1 1 -1 -1 1 -1 1 -1 1 1 1 1 1 -1 -1 1 -1 1 -1 1 -1 1 1 1 1 -1 1 1 1 -1 1 -1 -1 -1 1 1 1 1 1 -1 1 1 1 -1 1 1 -1 1 1 -1 1 1 -1 -1 -1 1 1 1 1 -1 1 1 -1 1 -1 -1 1 1 1 -1 1 -1 -1 1 1 1 1 1 -1 1 -1 1 -1 1 1 1 -1 1 -1 -1 1 -1 -1 1 -1 -1 -1 -1 1 -1 -1 1 1 1 -1 -1 -1 -1 1 -1 -1 1 1 -1 -1 -1 -1 -1 1 1 1 -1 -1 -1 1 1 -1 -1 1 -1 1 1 1 -1 -1 -1 1 1 1 1 1 -1 -1 1 -1 -1 1 -1 1 1 1 -1 -1 -1 1 -1 1 -1 1 -1 -1 1 1 1 1 1 -1 -1 1 -1 -1 -1 -1 1 -1 1 -1 1 -1 1 1 1 1 -1 1 1 -1 1 -1 -1 -1 -1 -1 1 -1 1 1 1 1 -1 1 1 -1 -1 -1 1 1 1 -1 -1 1 -1 -1 1 1 -1 1 -1 1 1 1 -1 1 1 1 -1 -1 -1 1 1 -1 -1 -1 1 1 1 1 -1 1 1 1 -1 1 -1 -1 -1 -1 -1 -1 -1 -1 -1 1 -1 -1 -1 -1 1 -1 -1 1 1 1 -1 1 1 -1 -1 -1 -1 -1 1 -1 1 -1 1 1 -1 -1 1 -1 1

$\bullet$ Solution of instance 37 with cut 7690:

-1 1 -1 1 -1 -1 1 -1 -1 -1 -1 -1 1 1 1 -1 -1 -1 1 1 -1 1 1 1 -1 -1 1 1 1 1 -1 -1 1 -1 1 1 -1 -1 -1 1 -1 -1 1 1 1 1 -1 1 1 1 -1 1 1 -1 1 1 1 -1 -1 -1 -1 -1 -1 -1 1 1 1 -1 -1 1 -1 1 -1 -1 1 -1 -1 1 1 1 1 -1 1 1 1 1 -1 1 -1 -1 -1 -1 -1 1 1 1 -1 1 -1 -1 -1 1 1 -1 -1 1 -1 1 -1 1 -1 -1 1 1 1 1 -1 1 -1 1 1 1 1 1 -1 1 -1 1 1 -1 -1 -1 -1 1 1 1 -1 1 1 1 -1 1 -1 -1 1 1 1 1 1 -1 -1 -1 1 1 -1 -1 -1 -1 1 -1 1 -1 1 1 -1 1 -1 -1 -1 1 1 1 -1 1 1 -1 1 -1 -1 1 1 1 1 1 1 1 -1 1 1 -1 -1 -1 1 1 1 1 -1 1 -1 1 -1 -1 -1 1 -1 -1 1 -1 -1 -1 -1 1 1 -1 -1 -1 1 1 1 1 -1 -1 1 -1 1 1 1 -1 1 -1 -1 -1 1 -1 -1 1 -1 -1 1 -1 -1 -1 -1 -1 -1 1 -1 1 1 -1 -1 -1 1 -1 1 1 -1 -1 1 1 1 1 1 1 -1 -1 -1 1 -1 -1 1 1 -1 1 -1 -1 -1 1 1 1 1 1 1 1 -1 -1 -1 -1 -1 -1 1 -1 1 1 1 -1 1 -1 -1 -1 -1 -1 -1 -1 1 -1 -1 -1 1 1 1 1 -1 1 -1 -1 1 1 1 -1 1 1 1 -1 -1 1 -1 -1 1 1 -1 -1 -1 1 -1 -1 1 -1 1 1 -1 -1 -1 -1 1 1 -1 1 -1 1 -1 -1 -1 1 1 1 -1 1 -1 -1 1 -1 -1 -1 1 1 -1 -1 -1 1 1 -1 -1 1 1 1 1 1 -1 1 1 1 1 1 1 -1 1 -1 1 -1 1 -1 1 1 -1 1 1 1 -1 1 1 -1 1 -1 1 -1 1 -1 1 -1 1 -1 -1 1 -1 -1 -1 -1 -1 1 -1 -1 1 1 1 1 1 -1 1 -1 -1 1 1 1 1 1 -1 1 1 1 1 -1 -1 -1 1 -1 -1 -1 1 1 1 -1 1 1 -1 -1 -1 -1 1 1 -1 1 -1 -1 1 -1 1 -1 -1 1 -1 1 -1 1 -1 1 1 -1 -1 -1 1 1 1 -1 -1 1 1 1 1 1 -1 1 -1 -1 -1 -1 -1 -1 -1 -1 1 -1 -1 -1 -1 -1 1 1 1 1 1 1 -1 1 -1 1 -1 -1 -1 -1 1 -1 -1 -1 -1 1 -1 1 1 -1 1 -1 1 -1 -1 1 -1 1 1 -1 -1 1 1 -1 1 1 -1 1 -1 1 -1 -1 -1 -1 1 1 -1 1 1 -1 1 -1 -1 -1 1 -1 1 1 -1 1 -1 1 -1 1 1 1 1 -1 -1 1 1 -1 -1 1 -1 1 -1 1 1 1 -1 1 -1 -1 1 -1 1 1 -1 -1 1 1 -1 -1 1 1 1 1 -1 1 1 -1 1 1 1 1 1 1 -1 -1 1 1 -1 1 1 1 1 1 1 -1 1 -1 1 1 1 1 -1 -1 -1 -1 -1 -1 1 1 -1 -1 -1 1 -1 -1 -1 1 -1 -1 -1 1 1 1 1 -1 1 -1 1 1 -1 1 -1 -1 1 1 -1 1 -1 -1 1 1 1 -1 -1 1 1 1 1 -1 -1 -1 1 -1 1 -1 1 -1 -1 -1 1 1 1 1 -1 1 1 -1 -1 1 -1 -1 1 -1 -1 -1 -1 -1 1 1 1 -1 -1 1 1 1 -1 1 1 1 1 1 1 -1 -1 1 -1 1 1 1 1 1 -1 1 1 1 -1 1 1 -1 1 -1 -1 -1 1 -1 1 -1 -1 -1 -1 1 -1 -1 -1 -1 1 1 1 -1 -1 1 1 1 -1 -1 1 1 1 -1 -1 1 1 -1 1 -1 -1 -1 1 -1 1 -1 -1 -1 1 -1 -1 -1 1 1 -1 -1 -1 1 1 -1 1 -1 -1 1 -1 1 1 1 1 -1 1 -1 -1 -1 1 -1 1 -1 1 1 1 -1 1 -1 -1 -1 1 -1 -1 1 -1 1 -1 -1 1 -1 1 1 -1 1 1 1 -1 1 1 1 -1 1 1 -1 1 -1 1 1 -1 1 1 1 1 -1 -1 -1 1 -1 1 -1 -1 -1 -1 -1 -1 -1 1 1 1 1 1 1 -1 1 -1 1 -1 1 -1 -1 1 1 -1 -1 1 1 -1 1 1 -1 1 -1 1 1 1 1 1 1 -1 1 -1 -1 1 1 -1 -1 1 1 -1 -1 1 1 1 1 1 -1 1 -1 -1 -1 1 -1 -1 -1 1 -1 1 -1 1 1 1 1 -1 1 -1 1 1 1 -1 -1 -1 1 -1 -1 -1 1 1 -1 1 1 1 -1 -1 -1 -1 -1 -1 -1 -1 1 -1 1 1 1 -1 1 -1 1 1 -1 -1 1 1 -1 1 1 -1 -1 -1 -1 -1 1 1 1 -1 1 1 1 1 1 -1 -1 -1 1 1 1 1 -1 -1 1 1 -1 -1 1 1 1 -1 1 1 -1 -1 -1 -1 1 1 1 1 1 -1 1 1 1 1 -1 1 -1 1 -1 1 -1 1 1 1 -1 -1 1 1 -1 -1 -1 1 1 1 -1 -1 1 1 1 1 -1 1 1 1 -1 -1 1 1 1 1 -1 -1 -1 -1 -1 1 1 -1 1 1 1 -1 -1 -1 -1 -1 -1 -1 -1 -1 -1 1 1 1 1 -1 -1 -1 1 -1 1 1 -1 1 -1 -1 -1 1 1 -1 -1 1 1 1 1 1 1 1 1 -1 -1 -1 -1 -1 -1 1 -1 1 -1 1 1 -1 1 -1 1 1 -1 -1 -1 1 -1 -1 1 -1 1 -1 -1 1 1 1 1 1 -1 1 -1 -1 -1 -1 1 1 -1 -1 1 -1 -1 1 -1 1 1 1 1 -1 1 1 -1 1 -1 -1 1 -1 -1 -1 1 1 -1 -1 1 1 -1 1 1 -1 -1 1 1 1 -1 1 -1 1 1 -1 1 1 -1 -1 -1 -1 1 1 1 -1 1 1 -1 1 1 -1 -1 1 -1 -1 1 -1 -1 -1 -1 1 1 1 -1 -1 -1 1 1 -1 -1 -1 1 1 -1 -1 1 -1 1 1 1 1 1 1 1 -1 -1 -1 1 1 1 -1 -1 1 1 1 -1 1 1 -1 1 1 -1 -1 -1 1 1 -1 -1 1 1 1 1 1 1 1 1 -1 -1 1 1 -1 1 -1 -1 1 1 1 -1 1 -1 -1 -1 1 1 -1 -1 1 -1 1 -1 -1 1 1 -1 1 -1 1 1 1 -1 1 1 -1 1 1 -1 -1 -1 -1 -1 -1 1 -1 1 -1 1 -1 1 1 1 -1 1 1 1 -1 1 1 -1 -1 1 -1 -1 -1 -1 1 -1 1 -1 1 -1 -1 1 1 -1 -1 -1 1 1 1 -1 -1 1 1 -1 -1 -1 -1 -1 -1 -1 1 -1 -1 -1 1 1 1 1 1 -1 1 1 1 1 -1 1 1 -1 1 -1 -1 -1 1 1 -1 -1 1 1 1 -1 -1 1 -1 1 -1 1 1 1 -1 1 -1 1 1 1 -1 -1 -1 1 1 1 1 -1 -1 -1 -1 1 1 1 1 1 1 -1 -1 -1 -1 1 -1 -1 1 -1 1 -1 -1 -1 -1 1 1 1 -1 -1 -1 1 -1 1 -1 1 1 -1 -1 1 -1 -1 1 -1 -1 1 -1 1 -1 -1 1 1 -1 -1 1 -1 1 1 -1 1 1 -1 -1 1 -1 -1 -1 -1 -1 -1 1 1 1 1 -1 -1 1 1 1 -1 1 1 -1 1 1 1 1 1 -1 1 1 -1 -1 1 1 1 1 -1 -1 1 -1 1 -1 -1 1 1 1 1 1 -1 -1 1 1 -1 1 1 -1 1 -1 1 1 1 1 1 -1 -1 -1 -1 1 -1 -1 -1 1 1 1 -1 1 -1 1 1 -1 1 -1 -1 1 -1 1 1 -1 1 -1 1 -1 1 -1 -1 -1 -1 -1 1 1 -1 -1 -1 -1 -1 1 -1 -1 -1 -1 1 1 1 1 1 1 -1 1 -1 1 1 -1 1 1 -1 1 -1 -1 1 1 1 -1 -1 -1 1 -1 1 -1 1 -1 1 -1 -1 -1 -1 1 1 -1 -1 1 1 -1 1 -1 -1 1 1 -1 -1 1 -1 1 1 -1 1 1 -1 1 -1 -1 1 -1 -1 1 1 -1 -1 1 1 -1 1 -1 1 -1 -1 1 1 1 1 1 1 1 1 1 1 1 1 1 -1 -1 -1 -1 -1 -1 1 1 1 -1 1 1 -1 -1 -1 -1 1 1 1 -1 -1 1 -1 -1 1 -1 -1 1 1 -1 1 -1 1 -1 -1 -1 -1 -1 1 -1 1 1 -1 1 1 1 -1 1 -1 -1 1 1 1 1 -1 -1 -1 -1 1 1 -1 1 -1 -1 1 -1 1 -1 1 -1 1 1 1 -1 1 1 -1 -1 -1 1 1 -1 1 -1 -1 -1 -1 1 -1 -1 1 -1 1 -1 -1 -1 1 -1 1 1 -1 -1 1 1 1 1 1 -1 -1 -1 1 -1 -1 -1 1 1 -1 1 1 -1 -1 1 1 1 -1 -1 1 1 -1 -1 -1 -1 1 1 -1 -1 -1 1 1 1 1 1 1 -1 1 -1 1 1 1 1 -1 -1 1 1 -1 1 -1 -1 1 -1 1 -1 -1 1 -1 1 1 1 1 1 -1 -1 -1 -1 -1 1 1 1 1 1 -1 1 -1 -1 1 1 -1 -1 1 1 -1 1 -1 -1 -1 1 1 1 1 1 1 1 1 1 1 1 -1 -1 -1 1 -1 -1 1 -1 1 -1 -1 1 -1 -1 -1 -1 1 -1 -1 -1 -1 -1 -1 -1 1 1 -1 -1 -1 1 1 -1 -1 1 1 1 1 -1 1 -1 -1 1 1 1 -1 -1 -1 -1 1 1 1 1 -1 -1 -1 -1 1 1 -1 -1 1 -1 1 -1 1 1 1 -1 -1 1 1 1 -1 1 1 -1 -1 1 -1 1 -1 1 1 1 1 -1 1 -1 1 1 1 1 -1 -1 1 -1 -1 1 1 1 -1 1 -1 -1 -1 1 -1 1 1 -1 1 1 1

$\bullet$ Solution of instance 38 with cut 7688:

-1 -1 1 -1 1 -1 1 1 -1 1 1 -1 1 -1 -1 1 1 -1 1 1 -1 1 1 -1 1 -1 1 1 -1 -1 -1 -1 1 -1 1 -1 1 -1 1 1 1 1 1 -1 1 1 -1 1 -1 -1 -1 1 -1 1 -1 1 1 1 -1 -1 -1 -1 -1 1 -1 1 -1 1 1 -1 1 1 1 1 -1 -1 1 -1 1 1 1 1 -1 -1 1 1 1 1 1 1 -1 1 -1 -1 1 1 1 -1 1 1 -1 -1 1 1 1 -1 1 -1 -1 1 -1 -1 1 1 1 -1 -1 -1 -1 1 -1 1 -1 -1 1 -1 1 -1 -1 1 1 1 1 -1 -1 -1 1 1 -1 1 -1 -1 1 1 1 1 1 1 -1 -1 1 -1 -1 1 1 -1 -1 1 -1 1 1 1 1 1 1 -1 1 -1 -1 1 -1 1 1 1 1 -1 1 -1 1 1 1 1 1 -1 -1 1 1 1 -1 -1 1 -1 1 1 -1 1 -1 -1 1 1 -1 -1 -1 1 1 -1 -1 -1 1 -1 -1 -1 -1 1 1 -1 1 1 1 -1 -1 1 1 -1 1 -1 1 -1 -1 -1 -1 1 -1 1 -1 -1 -1 -1 -1 1 1 -1 -1 1 1 -1 -1 -1 -1 -1 1 1 -1 1 -1 -1 1 1 1 -1 -1 1 -1 1 1 -1 1 -1 1 -1 -1 -1 1 1 1 1 1 1 1 -1 -1 -1 1 -1 1 1 -1 1 1 -1 -1 -1 -1 1 1 -1 -1 -1 -1 1 1 1 -1 -1 1 1 -1 -1 -1 1 1 1 1 -1 -1 -1 -1 1 1 1 1 1 1 -1 -1 1 -1 -1 -1 1 1 -1 -1 -1 -1 -1 -1 1 -1 1 -1 -1 1 1 1 1 1 -1 -1 -1 1 -1 1 1 1 1 1 -1 1 1 -1 -1 -1 -1 -1 1 1 1 1 -1 1 1 1 1 -1 -1 -1 1 1 -1 -1 1 1 1 -1 1 -1 1 -1 1 1 -1 -1 -1 -1 -1 1 1 -1 -1 1 -1 -1 1 -1 -1 1 -1 1 -1 1 1 -1 -1 -1 1 -1 1 -1 1 -1 -1 -1 1 1 -1 -1 -1 1 -1 1 -1 -1 -1 1 1 1 1 -1 1 -1 -1 -1 -1 -1 -1 1 1 -1 -1 1 1 -1 1 1 -1 -1 1 1 -1 -1 -1 -1 1 -1 1 -1 -1 -1 -1 1 -1 1 -1 -1 -1 -1 -1 -1 1 -1 -1 -1 -1 -1 1 -1 -1 -1 1 1 1 1 -1 1 1 1 1 -1 1 -1 -1 1 1 -1 1 -1 1 1 1 1 -1 -1 -1 -1 -1 1 -1 -1 -1 -1 1 -1 -1 -1 -1 -1 -1 -1 1 -1 1 -1 1 -1 1 1 1 1 1 1 1 -1 -1 1 1 1 -1 1 1 -1 -1 -1 -1 -1 1 -1 -1 1 -1 -1 -1 -1 1 -1 1 -1 -1 1 -1 1 1 -1 1 -1 1 -1 1 -1 1 1 -1 -1 -1 -1 -1 -1 -1 -1 1 -1 1 1 -1 -1 -1 -1 1 -1 -1 -1 -1 -1 1 -1 -1 1 1 -1 -1 1 -1 1 -1 1 -1 -1 -1 1 -1 -1 -1 -1 -1 1 -1 -1 1 1 -1 1 1 -1 -1 1 -1 1 -1 1 1 1 -1 -1 -1 1 1 -1 1 1 1 -1 -1 -1 -1 -1 1 1 1 1 1 1 1 -1 -1 -1 -1 1 -1 -1 -1 -1 -1 -1 -1 1 -1 -1 -1 -1 -1 -1 1 1 -1 1 1 1 -1 -1 1 1 1 1 1 1 -1 -1 -1 -1 -1 -1 -1 1 -1 -1 -1 -1 -1 -1 1 -1 1 -1 1 1 -1 -1 -1 -1 1 -1 -1 -1 1 -1 -1 1 -1 1 1 -1 -1 -1 -1 -1 -1 -1 1 1 1 -1 -1 -1 1 1 -1 1 1 -1 1 1 -1 -1 -1 1 -1 -1 -1 -1 -1 -1 1 -1 -1 1 -1 -1 -1 -1 1 -1 1 1 -1 1 1 1 1 1 -1 1 1 -1 -1 -1 -1 1 -1 1 -1 1 1 1 1 1 -1 -1 1 1 -1 1 -1 -1 -1 -1 -1 1 -1 -1 1 1 1 -1 -1 -1 1 1 1 -1 1 -1 -1 -1 -1 1 -1 -1 -1 1 1 -1 -1 1 -1 1 1 -1 1 -1 1 -1 -1 1 1 1 1 1 1 1 -1 -1 1 1 1 1 1 1 1 1 1 -1 -1 -1 -1 1 1 -1 1 1 1 -1 1 -1 -1 -1 1 -1 1 1 1 -1 -1 -1 1 1 1 1 -1 1 -1 1 -1 -1 1 1 1 1 1 -1 1 1 -1 -1 -1 -1 1 -1 1 -1 -1 -1 -1 1 -1 1 -1 -1 -1 -1 1 -1 -1 -1 1 -1 -1 1 -1 -1 -1 -1 1 -1 1 -1 -1 1 1 1 1 1 1 1 -1 -1 1 -1 -1 -1 -1 1 -1 1 -1 1 -1 -1 1 -1 1 1 -1 1 -1 1 1 -1 1 1 1 1 1 -1 1 -1 1 1 -1 1 1 -1 1 -1 -1 -1 -1 1 -1 -1 -1 -1 1 -1 1 -1 1 1 -1 -1 -1 1 -1 -1 1 1 1 -1 -1 -1 -1 1 -1 1 1 1 -1 -1 -1 1 1 -1 -1 1 -1 -1 1 -1 1 1 1 1 1 1 1 1 -1 -1 1 -1 -1 -1 -1 1 1 -1 1 1 1 -1 -1 -1 -1 1 1 -1 1 -1 1 -1 -1 1 1 1 1 1 -1 -1 1 1 -1 -1 1 -1 -1 1 -1 -1 1 1 -1 -1 1 -1 1 1 -1 1 -1 1 1 -1 1 -1 -1 1 -1 -1 1 1 -1 1 1 1 -1 1 1 -1 -1 -1 -1 -1 -1 1 1 1 1 -1 -1 -1 -1 -1 -1 -1 1 1 1 1 1 1 -1 1 1 -1 1 -1 1 1 1 1 1 1 -1 -1 -1 -1 -1 -1 1 1 -1 -1 -1 1 1 1 1 1 1 -1 1 -1 1 -1 1 1 -1 1 -1 1 1 -1 -1 -1 -1 1 1 1 -1 -1 1 -1 1 1 1 -1 1 -1 -1 1 -1 -1 -1 -1 1 1 1 -1 1 -1 1 -1 1 -1 1 -1 1 1 -1 -1 1 -1 -1 1 -1 1 -1 -1 -1 -1 1 -1 -1 1 1 -1 -1 -1 -1 1 -1 1 -1 1 1 -1 -1 -1 1 1 -1 1 1 -1 1 -1 -1 1 1 1 -1 -1 1 -1 1 -1 -1 1 1 1 1 1 1 1 -1 1 -1 -1 1 -1 1 -1 -1 -1 -1 -1 1 1 1 1 -1 -1 -1 1 1 1 -1 -1 1 -1 1 1 1 1 1 1 -1 1 1 -1 -1 -1 1 1 -1 -1 1 -1 1 -1 1 1 1 -1 -1 -1 1 1 1 -1 -1 1 -1 -1 -1 1 1 1 1 1 -1 1 1 -1 1 -1 -1 -1 -1 -1 1 1 -1 -1 1 1 1 -1 -1 1 1 1 1 -1 -1 1 1 -1 1 -1 1 1 1 1 1 1 -1 -1 1 -1 -1 -1 -1 1 -1 -1 1 -1 -1 -1 1 -1 1 1 -1 -1 -1 -1 -1 -1 -1 -1 1 -1 1 -1 1 -1 -1 -1 -1 1 -1 -1 1 -1 1 1 -1 1 -1 1 -1 -1 1 -1 1 1 -1 1 -1 1 1 1 1 1 -1 -1 1 -1 1 1 -1 -1 1 1 1 1 1 1 -1 1 -1 1 -1 1 -1 -1 1 -1 -1 -1 1 -1 -1 -1 -1 -1 -1 -1 1 1 -1 -1 1 1 -1 -1 -1 1 -1 1 -1 1 -1 1 1 1 1 1 1 -1 -1 1 1 -1 1 1 1 -1 -1 1 1 -1 1 -1 1 1 -1 1 -1 1 -1 -1 -1 1 -1 1 -1 1 1 -1 1 -1 1 -1 -1 -1 1 -1 1 -1 -1 1 1 1 -1 -1 1 -1 1 -1 1 -1 -1 1 -1 1 1 1 1 -1 1 -1 1 -1 1 -1 -1 1 -1 -1 -1 -1 1 -1 1 1 -1 -1 1 -1 1 -1 1 1 1 -1 1 1 -1 1 -1 1 -1 1 1 -1 -1 1 -1 1 1 1 -1 1 -1 1 1 -1 1 -1 1 1 1 1 1 1 -1 -1 1 1 -1 -1 1 1 -1 -1 1 -1 -1 -1 1 1 -1 -1 1 1 1 1 1 -1 1 -1 -1 -1 1 -1 1 1 1 1 -1 1 -1 1 -1 -1 1 1 1 1 1 1 -1 -1 1 -1 1 1 -1 -1 1 1 1 -1 1 -1 1 1 1 1 -1 1 1 -1 1 1 -1 1 -1 -1 1 -1 -1 1 1 -1 -1 1 -1 1 1 -1 -1 -1 1 1 1 1 1 -1 1 -1 1 1 1 1 1 1 -1 -1 -1 -1 -1 1 -1 -1 1 1 -1 1 -1 1 1 1 -1 1 1 -1 1 1 -1 1 -1 -1 1 1 1 1 1 1 1 1 1 -1 -1 1 -1 1 -1 1 1 -1 -1 -1 -1 -1 1 -1 -1 1 1 1 -1 1 1 -1 -1 1 1 -1 -1 -1 -1 -1 -1 1 -1 1 1 -1 1 1 1 -1 -1 -1 1 -1 1 -1 1 -1 -1 1 1 1 1 1 -1 1 1 -1 -1 -1 -1 -1 1 1 1 1 -1 1 1 1 1 -1 -1 -1 1 -1 -1 1 1 -1 1 1 -1 -1 1 -1 -1 1 1 1 1 1 1 1 -1 1 -1 -1 -1 1 -1 -1 -1 -1 -1 -1 -1 1 -1 1 1 1 -1 1 -1 1 -1 1 -1 1 -1 1 1 -1 -1 1 -1 1 1 1 -1 -1 -1 1 -1 1 1 1 -1 -1 1 -1 1 1 -1 -1 1 1 1 -1 1 -1 -1 1 1 -1 1 -1 1 1 1 1 -1 -1 1 1 -1 1 -1 -1 1 -1 1 -1 1 1 -1 -1 -1 -1 -1 -1 1 1 1 1 1 1 1 -1 1 -1 -1 1 -1 1 -1 1 -1 -1 1 -1 -1 1 -1 1 -1 -1 1 1 1 1 1 1 1 -1 1 1 1 -1 -1 -1 1 -1 1 1 1 1 1 -1 1 -1 -1 -1 1 1 -1 1 -1 -1 -1 -1 -1 -1 1 -1 1 -1 1 1 1 1 1 1 -1 1 1 -1 1 1 1 1 1 1 -1

\end{document}